\documentclass[11pt]{article}

\usepackage{amsmath}
\usepackage{epsfig}
\usepackage{graphics}

\newcommand{\N}{N\raise.7ex\hbox{\underline{$\circ $}}$\;$}

\begin{document}

\title{
 V.M. Red'kov\footnote{redkov@dragon.bas-net.by}\\
 Spinor Structure of   $P$-Oriented Space,
Kustaanheimo-Stifel \\ and Hopf Bundle  -- Connection between
Formalisms \\
{\small Institute of Physics, National  Academy of Sciences of Belarus
 } }

\maketitle

\begin{quotation}

 In the work some  relations  between three techniques,
Hopf's bundle, Kustaan\-heimo-Stiefel's
 bundle,  3-space with spinor structure have been examined. The  spinor space is viewed
as a real space that is minimally (twice as much) extended in
comparison with an ordinary vector 3-space:  at this instead of
$2\pi$-rotation now only $4\pi$-rotation is taken to be the
identity transformation in  the geometrical space. With respect to
a given $P$-orientation of an initial unextended  manyfold, vector
or pseudovector one, there may be constructed two different
spatial spinors, $\xi$ and $\eta$, respectively. By definition,
those spinors provide us  with points of the extended  space
models, each spinor is in the correspondence $2 \longrightarrow 1$
with points of a vector space. For  both models an explicit
parametrization of the spinors  $\xi$ and   $\eta$ by spherical
and parabolic coordinates is given, the parabolic system turns out
to be the  most convenient  for simple defining spacial spinors.
Fours of real-valued coordinates by Kustaanheimo-Stiefel, $U_{a}$
and $V_{a}$, real and  imaginary parts of complex spinors $\xi$
and   $\eta$ respectively, obey two quadratic constraints. So that
in both cases, there exists a Hopf's mapping from the part of
3-sphere  $S_{3}$ into the entire 2-sphere $ S_{2}$. Relation
between two spacial spinor is found: $ \eta =  ( \xi - i \; \sigma
^{2} \xi ^{*} )/\sqrt{2} $, in terms of Kustaanheimo-Stiefel
variables  $U_{a}$ and $V_{a}$ it is a  linear transformation from
$SO(4.R)$,  which does not enter its sub-group generated by
$SU(2)$-rotation over spinors.

\end{quotation}


\section{Introduction}

\hspace{5mm} In the  literature  [1-66] there exist three
terminologically different, but close in mathematical technique,
approaches. These are spinor space structure\footnote{See for example
[29,61-66], however an idea itself to employ  in all parts exceptionally spinors
in place  of vectors may be seen in the very creating the  quantum theory; see,
 in particular, Cartan's and Weyl's works  [2-4].},
 Hopf's bundle [5],  Kustaanheimo-Stifel bundle  [9,10].

It is  interesting to note that before  creating the Cartan's  group theory of spinors
Darboux' lectures  [1]  contain description of a map which
in  present-day  terminology can be referred to a spinor representation for spinors,
or in other words to  Hopf's  and  Kustaanheimo-Stifel formalism.

Differences between three mentioned approaches consist mainly in conceptual  accents.

In the Hopf's technique it is suggested to use in all parts only complex spinors $\xi$
and conjugate $\xi^{*}$ instead of real-valued vector (tensor) quantities.
This approach is used mainly for a vector that can be associated
with  a pseudo vector space model $(x,y,z)$; at this a phase factor at $\xi$ does not influence
real-valued vector coordinates $(x,y,z)$.

In the Kustaanheimo-Stifel  approach we are to use four real-valued coordinates,
from which  by means of definite bi-linear functions Cartesian coordinates $(x,y,z)$
can be formed up. These Kustaanheimo-Stifel's  four variables
are real and imaginary parts of two spinor components.
The  known spinor invariant under $SU(2)$,  being transformed to these variables,
becomes the sum of four squared quantities, so that  we can associate spinor
technique with geometry of 3-dimensional Riemann space $S_{3}$.

In essence, the Kustaanheimo-Stifel's approach is elaboration of complex spinor,
 $\xi$ and $\xi^{*}$,  Hopf's technique in term  of four real-valued variables.
 At this we  are able to hide (in appearance only)  the  presence  in  the
 formalism of  the  non-analytical operation of  complex conjugation.

Spinor space structure, formalism developed in present work, also
exploits possibilities given by spinors to construct  3-vectors, however  the
emphasis is taken to doubling the set of spatial points so that  we get an extended
space model that is called  a space with spinor structure [61-66].
In such an extended  space, in place of  $2\pi$-rotation,  only  $4\pi$-rotation
transfers the space into itself.
In the present work this extended  space model is investigated with the  help
of curvilinear coordinates [66]. At this   we consistently  distinct
two sorts of spatial spinors: which are referred to  a vector and pseudo vector
models respectively. Else one point; as shown, the  procedure itself of doubling the  manyfold
can be  realized easier when for parameterizing  the space some curvilinear
coordinate system is used in contrast with the Cartesian coordinates.
In the work we consider in detail two coordinates systems: spherical
 and parabolic ones.

It should be  noted that because all three technique exploit much the same mathematical
formalism, always it is possible transportation results from one technique to another.

The main  purpose of the present work is to describe some
relations between these three approaches. In particular, it is
shown the following: the structure of spinor space model can  be
considered as minimization of  (Hopf and Kustaanheimo-Stifel)
mapping   in the sense that we can eliminate from 3-dimension set
$U_{1}^{2}  +U_{2}^{2} + U_{3}^{2} + U_{4}^{2} = 2r $, Riemann
sphere  $S_{3}$,   all the  redundant points so that the remaining
2-dimension set  $\bar{S}_{3} \in S_{3}$  is mappable into
2-sphere $x^{2} + y^{2} + z^{2} = r^{2}$ by taking the rule  $2
\longrightarrow 1$. In addition, different spinor space model are
established in dependence of $P$-orientation of  an initial vector
space.

\section{
Spinor extension of a  pseudo vector space model
 }

\hspace{5mm} Extending  the  vector space with the aim to describe
its spinor structure may be done  on the base of the known
decomposition of two rank spinor (we  mean spinor under  Cartan's
extended unitary group  $SU(2)$
[1-3]
\begin{eqnarray}
\xi \otimes \xi^{*} = ( r + x_{j} \sigma^{j})\; , \qquad
\label{2.1a}
\\
x_{j} = {
1\over 2} \; \mbox{sp} \;[ \sigma^{j} (\xi \otimes \xi ^{*})] = {1 \over 2} \;
\xi^{+} \sigma^{i} \xi \; ,
\nonumber
\\
 r = {1 \over 2} \; \mbox{sp} \;
(\xi \otimes \xi ^{*}) =
{1 \over  2}  \; \xi^{+} \xi =
\mbox{inv} \; .
\nonumber
\end{eqnarray}

Under continuous transformations from  $SU(2)$ group the coefficients
 $x_{j}$ in  (\ref{2.1a}) transform as 3-vector representation
[60]
\begin{eqnarray}
 B(n) = I \; n_{4} - i \;\sigma^{j} n_{j} \; , \;
 \nonumber
 \\
n_{4}^{2} + n_{j}n_{j} = 1 \; , \;
 \xi ' = B(n) \; \xi \; ,
 \nonumber
\\
 O (n)  =  I + 2 \; [ \; n_{4} \;\vec{n}\; ^{\times} + ( \vec{n}\; ^{\times})^{2} \; ]\;  ,
\nonumber
\\
x_{k}' = O_{kl} (n) \;  x_{l} \; , \;
\qquad  ( \vec{n}\; ^{\times})_{kl}= -\epsilon _{klj} \;n _{j} \; .
\nonumber
\end{eqnarray}

\noindent
Indeed,
\begin{eqnarray}
x'_{k} =  {1 \over 2} \;
\xi ^{'+} \sigma^{k} \xi' = {1 \over 2} \; \xi ^{+}   [
 B^{+}(n)  \sigma^{k} B (n) ] \;  \xi
= {1 \over 2} \; \xi ^{+} \;O_{kl}(n)  \sigma ^{l} \; \xi = O_{kl}
(n) \; x_{l} \; . \label{2.1b}
\end{eqnarray}

\noindent
Under Cartan's spinor $P$-reflection, at any intrinsic parity  $\delta = \pm 1$  of spinor
$\xi$, the  quantity  $x_{j}$ transforms as a pseudo vector:
\begin{eqnarray}
\xi ' = \delta  \; J \; \xi \;, \; J = \left | \begin{array}{cc} i & 0 \\0 & i
\end{array} \right  | ,  \; x_{j}' = + x_{j} \; .
\label{2.1c}
\end{eqnarray}

Solving eq. (\ref{2.1a}) with respect to spinor  $\xi$
that parameterizes  an initial pseudo vector space in accordance with the rule
  $2 \longrightarrow 1$,  lead us to the  following formulas ( take notice  to $
\phi \in [ -2\pi , + 2 \pi ]   $)
\begin{eqnarray}
\xi (x_{1},x_{2},x_{3}) = \left | \begin{array}{c} U_{1} + i\;
U_{2}
\\ U_{3} + i \;U_{4}
\end{array} \right | =
\left | \begin{array}{lr}
\sqrt{r + x_{3} }\; & \; e^{-i\phi/2 }  \\
\sqrt{r -  x_{3}  } \;& \;  e^{+ i\phi/2 }
\end{array} \right |  \; ,
\nonumber
\\
r = \sqrt{x_{1} ^{2} + x_{2}^{2} + x_{3}^{2}} \; , \qquad
e^{i\phi} = {x_{1} + i \;x_{2} \over \sqrt{ x_{1}^{2} +
x_{2}^{2}}} \; , \; \label{2.2}
\end{eqnarray}

\noindent
real-valued $U_{1},...,U_{4}$  coincide with Kustaan\-heimo-Stifel variables.

As we  employ spherical coordinates, the  space spinor  $\xi$ will  take the form [66]
\begin{eqnarray}
\xi (r,\theta,\phi) = \left | \begin{array}{lr}
\sqrt{r\;( 1 + \cos  \theta) }\; & \; e^{-i\phi/2 }  \\
\sqrt{r\; (1 - \cos \theta ) } \;& \; \; e^{i\phi/2 }
\end{array} \right |  \; , \; \;
 \phi \in [ -2\pi , + 2 \pi ] \; .
\label{2.3}
\end{eqnarray}

\noindent  Thus, four  Kustaan\-heimo-Stifel coordinates $U_{a}$
are functions of three spherical ones $(r,\theta,\phi)$,
 so that we are to expect some additional relationship between these four parameters.

Further we  get to
\begin{eqnarray}
{1 \over 2} \; ( \;\xi^{+} \xi \;) =
{1 \over 2} \;  (  U_{1}^{2} +  U_{2}^{2} +
U_{3}^{2} +  U_{4}^{2}   ) = r \; ,
\label{2.4}
\\
{1 \over 2} \;( \;\xi^{+} \sigma^{1} \xi \; ) =
 U_{1}\; U_{3}  + U_{2} \;U_{4}
 =  r \; \sin \theta \; \cos \phi  = x _{1}\;  ,
\nonumber
\\
{1 \over 2} \;( \;\xi^{+} \sigma^{2} \xi \; ) =  U_{1} \; U_{4} -
U_{2} \; U_{3} =
 r \; \sin \theta\; \sin \phi  = x_{2} \;  ,
\nonumber
\\
{1 \over 2} \;( \; \xi^{+} \sigma^{3} \xi \; )   ={1 \over 2} \;
 ( U_{1}^{2} + U_{2}^{2} - U_{3}^{2} - U_{4}^{2} )
=  r \; \cos \theta   = x_{3}  \;  .  \label{2.5c}
\end{eqnarray}

In other words, Cartesian coordinates  $x_{1},x_{2},x_{3}$ have been expressed
as bi-linear combinations of variables $U_{1},...,U_{4}$. It should be  noted that
the real-valued  parameters $U_{a}$ taken from (\ref{2.2}) and   (\ref{2.3}) obey  the identity
\begin{eqnarray}
U_{1}\;  U_{4} + U_{2} \; U_{3} = 0 \; .
\label{2.6}
\end{eqnarray}

\noindent
which points out that we  have dealings with the Hopf mapping
$S_{3} \Longrightarrow S_{2}$ with the rule
$2 \longrightarrow 1$   from special sub-set in 3-sphere into the full 2-sphere:
\begin{eqnarray}
\{ \;  U_{a}  U_{a} = 2r, U_{1} U_{4} + U_{2} U_{3} = 0\;
 \} \qquad  \Longrightarrow
\qquad  \{ \; x_{1} ^{2} + x_{2}^{2} + x_{3}^{2} = r^{2} \;\} \; .
 \label{2.7}
\end{eqnarray}

\section{Spinor extension of a vector space model}

\hspace{5mm}
Constructing a spinor model for a vector space can be done in the same line   but on the
base of the  formula (decomposition of direct product of a spinor by  itself in terms
 of Pauli matrices)
\begin{eqnarray}
\eta \otimes \eta   = ( a_{j}  +i \;x_{j})\; \sigma^{j} \;
\sigma^{2}, \qquad (a_{j} + i \;x_{j} )= {1 \over 2} \; \mbox{sp}
\;  [\sigma^{2} \sigma^{j} (\eta \otimes \eta ) ] \; .
\label{3.1a}
\end{eqnarray}

\noindent  It is easier verified that $( a_{j}  +i\; x_{j})$ transforms
as a  3-vector under  $SU(2)$ group (its real and imaginary parts transform independently
by means of real-valued rotation matrix).
Indeed
\begin{eqnarray}
 a'_{j}  +i x'_{j}  =
 {1 \over 2}  \tilde{\eta}'  \sigma^{2} \sigma^{j}  \eta'
=  {1 \over 2}  \tilde{\eta}  \; [ \tilde{B}(n)
 \sigma^{2} \sigma^{j}  B(n)  ] \; \eta ,
\nonumber
\end{eqnarray}

\noindent from here, with the  known identity for spinor
transformations  [60]
\begin{eqnarray}
\tilde{B}(n)  \sigma^{2} = \sigma^{2} B^{-1}(n) \; , \qquad
B^{-1}(n) \sigma^{j}  B(n) = O_{jl}(n)  \sigma^{l} \; ,
\nonumber
\end{eqnarray}

\noindent we get
\begin{eqnarray}
( a'_{j}  +i\; x'_{j} ) = O_{jl}(n) \;  ( a_{l}  +i \;x_{l}) \; .
\nonumber
\end{eqnarray}

As seen, under Cartan's spinor  $P$ -- reflection (at any intrinsic parity
$\delta = \pm 1$  of the
spinor  $\eta$), the quantity  $(a_{j} + i\;x_{j})$ behaves as a vector:
\begin{eqnarray}
\eta ' = \delta  \; J \; \eta , \; J = \left | \begin{array}{cc} i & 0 \\0 & i
\end{array} \right  | , \qquad
 (a_{j}'+ i \; x_{j}') = - \; (a_{j} + i\;   x_{j} )\; .
\label{3.1b}
\end{eqnarray}

Solving eq.  (\ref{3.1a}) with respect to  $\eta$,
which  parameterizes the vector space   $x_{j}$
by scheme   $2 \longrightarrow 1$, we get to  [66]
\begin{eqnarray}
\eta^{\sigma} (x_{1},x_{2},x_{3}) =
\left | \begin{array}{c} V_{1}
+ i\;V_{2} \\ V_{3} + i\; V_{4}
\end{array} \right |
= \left | \begin{array}{lr}
\sqrt{x - \rho }\; & \; \sigma \; e^{-i\phi/2 }  \\
\sqrt{x +  \rho } \;& \; \; e^{ +i\phi/2 }
\end{array} \right |  \; ,
\nonumber
\\
\rho = \sqrt{x_{1}^{2} + x_{2}^{2}} , \qquad e^{i\phi} = {x_{1} +
i\; x_{2} \over \sqrt{ x_{1}^{2} + x_{2}^{2}}}, \qquad
 \phi \in [ -2\pi , + 2 \pi ] \; ;
\label{3.2}
\end{eqnarray}

\noindent where  $\sigma = +1$ corresponds to upper half-space   $x_{3}
>0$ , and $\sigma = -1$ corresponds
to the half-space  $x_{3} < 0$.

In particular, spinor  $\eta$  looks in spherical coordinates as
follows  [66]
 \begin{eqnarray}
\eta ^{\sigma} (r,\theta,\phi) = \left | \begin{array}{lr}
\sqrt{r\;( 1 - \sin  \theta) }\; &  \sigma \;e^{-i\phi/2 }  \\
\sqrt{r\; (1 + \sin  \theta ) } \;& \; e^{i\phi/2 }
\end{array} \right |   .
\label{3.3}
\end{eqnarray}

\noindent
This representation of the spinor can be  easily verified by direct
calculation. Indeed
\begin{eqnarray}
 {1 \over 2 } \; \mbox{sp} \; [\sigma^{1}
(\eta^{\sigma} \otimes \eta^{\sigma} ) \sigma^{2} ] = a_{1} + i
\;x_{1} = -r \sin \phi + i \; r \sin \theta \cos \phi \; ,
\nonumber
\\
 {1 \over 2 } \; \mbox{sp} \; [\sigma^{2}
(\eta^{\sigma} \otimes \eta^{\sigma} ) \sigma^{2} ] = a_{2} + i
\;x_{2} = r \cos \phi + i \; r \sin \theta \sin  \phi \; ,
\nonumber
\\
 {1 \over 2 } \; \mbox{sp} \; [ \sigma^{3}
(\eta^{\sigma} \otimes \eta^{\sigma} ) \sigma^{2} ] = a_{3} + i\;
x_{3} = 0 +   i \; r \cos  \theta  \; . \label{3.4c}
\end{eqnarray}

The  identity  $a_{3} = 0$ is not accidental, it takes place
at any coordinate system. Indeed, with the help of eq.
(\ref{3.2}) we get
\begin{eqnarray}
(\eta \otimes \eta) = \left | \begin{array}{cc}
(x-\rho) e^{-i\gamma} & \sqrt{x^{2} - \rho^{2}} \; \sigma \\
\sqrt{x^{2} - \rho^{2}} \; \sigma  & (x+\rho) e^{+i\gamma}
\end{array} \right |
= \left | \begin{array}{cc}
(x-\rho) e^{-i\gamma} &  x_{3} \\
x_{3}  & (x+\rho) e^{+i\gamma}
\end{array} \right | \;
\nonumber
\end{eqnarray}

\noindent   and  further
\begin{eqnarray}
a_{1} + i \;x_{1} = - x \sin \gamma + i \;\rho \cos \gamma \;,
\nonumber
\\
a_{2} + i \;x_{2} = + x \cos  \gamma + i \;\rho \sin \gamma \; ,
\nonumber
\\
a_{3} + i\; x_{3} =  0 + i \;\sqrt{x^{2} - \rho^{2}} \; \sigma   \; .
\label{3.5}
\end{eqnarray}

Let us produce the formulas connecting Cartesian coordinates
$x_{j}$  from (\ref{3.4c})   with corresponding
Kustaanheimo-Stifel variables. The  problem is reduced to
expressing  $a_{j} + i\; x_{j}$ in terms of real-valued parameters
$V_{1},...,V_{4}$:
\begin{eqnarray}
a_{1} + i \;x_{1}  \qquad \qquad \qquad
\label{3.6}
\\
=V_{1} V_{2} - V_{3} V_{4} +
i {-V_{1}^{2} + V_{2}^{2} + V_{3}^{2} - V_{4}^{2} \over 2} \;,
\nonumber
\\
a_{2} + i\; x_{2} \qquad \qquad \qquad
\nonumber
\\
=  {V_{1}^{2} - V_{2}^{2} + V_{3}^{2} -
V_{4}^{2} \over 2}  +  i  (V_{1} V_{2} + V_{3} V_{4})\; ,
\nonumber
\\
i x_{3} = -V_{1}   V_{4} - V_{2}  V_{3} + i (V_{1}
 V_{3} - V_{2} V_{4}) \; .
\nonumber
\end{eqnarray}

Additionally we can easily produce  the formula for $r^{2}$ in terms
$V_{1},...,V_{4}$:
\begin{eqnarray}
x_{1}^{2} + x_{2}^{2} + x_{3}^{2} = {1 \over 4} \; (V_{1}^{2}  +
V_{2}^{2} + V_{3}^{2} + V_{4}^{2}) ^{2} \; .
\label{3.7a}
\end{eqnarray}

\noindent This means that again we have  dealings with
the Hopf map from a party of the  sphere  $S_{3}$  into the full  2-sphere  $ S_{2}$:
\begin{eqnarray}
\{  \;   V_{a}  V_{a} = 2r\;, \; V_{1} \;  V_{4} + V_{2} \; V_{3}
=0 \;  \} \;\;   \Longrightarrow \qquad  \{ \; x_{1}^{2} +
x_{2}^{2} + x_{3}^{2} = r^{2} \;\} \; . \qquad \label{3.7b}
\end{eqnarray}

\section{ Parabolic coordinates  and determining
spatial spinors}

\hspace{5mm} It is turned out that the  known  parabolic
coordinates plays a special   role in defining spatial spinor
$\xi$  and  $\eta$. First let us  consider  the pseudo vector
model (see also in [66]). Let  the  spatial spinor  $\xi$ be given
by its natural parameters $(N,M,\phi)$:
\begin{eqnarray}
\xi  = \left | \begin{array}{c} N \; e^{-i\phi /2} \\
M \; e^{+i \phi /2 } \end{array} \right | \; , \; \phi \in [ -2\pi
, +2 \pi ]\; , \qquad
M \in [0 , + \infty ) , \qquad N \in [0 , +
\infty ) ,
 \; .
 \label{4.1}
 \end{eqnarray}

It is easily to show that $(N,M,\phi)$ coincides with the known
parabolic coordinates (to the case of  vector  model there corresponds
the domain $\gamma \in [0, + 2 \pi ]$).  To this end, with the  help of
\begin{eqnarray}
\xi \otimes \xi^{*} = \left | \begin{array}{cc}
N^{2} & NM \; e^{-i\phi} \\
NM \; e^{+i\phi} & M^{2}
\end{array} \right  |
\nonumber
\end{eqnarray}

\noindent
one  finds
\begin{eqnarray}
x_{1} = {1 \over 2} \; \mbox{sp} \; [ \sigma^{1} \xi \otimes \xi^{*} ]
 =
NM \cos \phi \; ,
\nonumber
\\
x_{2} = {1 \over 2} \;
\mbox{sp} \; [ \sigma^{2} \xi \otimes \xi^{*} ]
 =
NM \sin \phi \; ,
\nonumber
\\
x_{3} = {1 \over 2} \; \mbox{sp} \; [ \sigma^{3} \xi \otimes \xi^{*} ]
 = { N^{2} - M^{2} \over 2}  \; ,
\nonumber
\\
r=   {1 \over 2} \; \mbox{sp} \; [  \xi \otimes \xi^{*} ]
 =  { N^{2} + M^{2} \over 2}  \; ,
\label{4.2c}
\end{eqnarray}

\noindent
From this, taking into account determining relations for parabolic
coordinates
\begin{eqnarray}
x_{1} = \xi \; \eta \; \cos \phi, \; x_{2} = \xi \; \eta \; \sin
\phi, \; \; x_{3} ={ \xi^{2} - \eta^{2} \over 2} \; , \; r = {
\xi^{2} + \eta^{2} \over 2}\; , \label{4.3}
\end{eqnarray}

\noindent we may conclude that spinor  $\xi$ natural parameters
$(N,M,\phi)$ can be identified with
$ (\xi, \eta, \phi)$. The  variable   $U_{a}$  are given as by
\begin{eqnarray}
U_{1} = N \; \cos{ \phi\over 2} \;
 , \; U_{2} =  - N \; \sin {\phi \over 2}\; ,
 \;
 U_{3} = M \; \cos{ \phi\over 2} \;
 , \; U_{4} = N \; \sin {\phi \over 2} \; .
\label{4.4}
\end{eqnarray}

\noindent From this  it follows
\begin{eqnarray}
U_{1}\; U_{4} + U_{2} \;U_{3} = 0 \; .
\label{4.5}
\end{eqnarray}

Now let us proceed to the second model.
Let us designate parabolic coordinates as $(N,M,\phi)$. With the  help of
\begin{eqnarray}
x_{1} = NM \; \cos \phi , \qquad  x_{2} = NM \; \sin \phi,
 \nonumber
 \\
 x_{3} = { N^{2}  - M^{2} \over 2}\; , \;
 r = {N^{2} + M^{2} \over 2} , \; \rho  = N\; M \;  ,
\nonumber
\end{eqnarray}

\noindent
we get
\begin{eqnarray}
 (\;  \sigma \;)\; (+ \sqrt{x- \rho} \; )    =
 {N- M \over \sqrt{2}} \; , \qquad
 + \sqrt{x+ \rho}   =   {N+ M \over \sqrt{2 }} \; .
\label{4.6}
\end{eqnarray}

\noindent Therefore,  spatial spinor  $\eta$  looks as
(take notice that   $\sigma = \pm 1$  disappears)
\begin{eqnarray}
\eta =
{1 \over \sqrt{2}} \; \left | \begin{array}{c}
(N-M) \;e^{-i\phi/2} \\[2mm]
(N+M) \; e^{+i\phi/2}
\end{array} \right | \; .
\label{4.7}
\end{eqnarray}

\noindent  From this it follows
\begin{eqnarray}
a_{1} + i x_{1} = -  {N^{2} + M^{2} \over 2}  \sin \phi + i  NM  \cos \phi \; ,
\nonumber
\\
a_{2} + i x_{2} =
 +  {N^{2} + M^{2} \over 2}  \cos \phi + i  NM  \sin  \phi \; ,
\nonumber
\\
a_{3} + i x_{3} =  0 + i  {N^{2} - M^{2} \over 2} \; .\qquad \qquad
\label{4.8c}
\end{eqnarray}

The variables  $V_{1},...,V_{4}$ are given by
\begin{eqnarray}
V_{1} =  { N - M \over \sqrt{2}}  \cos {\phi \over 2} \; , \;
V_{2} =  - {  N -M \over \sqrt{2}}  \sin  {\phi \over 2}  \; ,
\nonumber
\\
V_{3} =  {N + M \over \sqrt{2}}  \cos {\phi \over 2} \; , \;
V_{4} =  + { N + M \over \sqrt{2}}  \sin  {\phi  \over 2} \; .
\label{4.9}
\end{eqnarray}

\section{Connection between the  variables  $U_{a}$ and  $V_{a}$ }

\hspace{5mm}
Comparing  (\ref{4.9}) with  (\ref{4.4}), we readily establish relationships:
\begin{eqnarray}
V_{1} = {   U_{1} - U_{3} \over \sqrt{2}}\; , \;
V_{2} = {   U_{2} + U_{4} \over \sqrt{2}} \; ,
\nonumber
\\
V_{3} = {U_{1} + U_{3} \over \sqrt{2}} \;, \;
V_{4} = {-U_{2} + U_{4} \over \sqrt{2}}\; ,
\label{5.1a}
\end{eqnarray}

\noindent
or in a matrix form
\begin{eqnarray}
\left | \begin{array}{c}
V_{4} \\ V_{1} \\ V_{2} \\ V_{3}
\end{array}  \right |
=
{1 \over \sqrt{2}}\; \left | \begin{array}{cccc}
1 & 0 & -1 & 0 \\
0 & 1 &  0 & -1 \\
1 & 0 &  1 & 0 \\
0 & 1 & 0  & 1
\end{array} \right |
\left |  \begin{array}{c}
U_{4} \\ U_{1} \\ U_{2} \\ U_{3}
\end{array}  \right | \; .
\label{5.1b}
\end{eqnarray}

Returning with (\ref{5.1b}) to  (\ref{3.2}), for  $\eta (V_{a})$
one  produces
$(U_{1})$:
\begin{eqnarray}
\eta  = \left | \begin{array}{c}
V_{1} + i\; V_{2} \\[2mm]
V_{3} + i\; V_{4}
\end{array} \right |
= {1 \over \sqrt{2}} \; \left | \begin{array}{c}
(U_{1} -U_{3})  + i\;(+U_{2}+U_{4}) \\[2mm]
(U_{1} +U_{3})  + i\;(-U_{2}+U_{4})
\end{array} \right | ,
\label{5.2a}
\end{eqnarray}

\noindent
with  the relations
\begin{eqnarray}
\left. \begin{array}{ll}
U_{a} U_{a} = 2r \; , &\;
V_{a} V_{a}  = 2r \; ,\\[2mm]
U_{1} \;U_{4} + U_{2} \;U_{3} = 0 \; ,  & \; V_{1}\; V_{4} + V_{2} \;V_{3} = 0 \; .
\end{array} \right.
\label{5.2b}
\end{eqnarray}

Therefore, an explicit form of two spatial spinors $\xi$ and  $\eta$ differs
when we use ordinary curvilinear coordinates (cartesian, spherical, parabolic)
 and also when we use Kustaanheimo-Stifel variables.
This means that just these explicit forms of two sorts  of spinors
describes  essential difference of two spinor space  models.

Relation between    $\xi$ and    $\eta$ can be  readily
expressed in terms of  complex spinor themselves.
Indeed, with the help of
\begin{eqnarray}
\eta  = {1 \over \sqrt{2}} \left | \begin{array}{c}
(N - M) \; e^{-i \phi /2}  \\
(N  + M) \; e^{+i \phi/2}
\end{array} \right|  ,
\xi  = \left | \begin{array}{c}
N \; e^{-i \phi /2}   \\   M \; e^{+i \phi /2}
\end{array} \right |.
\nonumber
\end{eqnarray}

\noindent one  produces
\begin{eqnarray}
\eta = {1 \over \sqrt{2}} ( \xi - i \; \sigma ^{2} \xi ^{*} ) \; . \qquad
\label{5.3a}
\end{eqnarray}

\noindent inverse relation looks as
\begin{eqnarray}
\xi = {1 \over \sqrt{2}} \; ( \eta \;-\; i \;\sigma ^{2} \eta ^{*} )
\; .
\label{5.3b}
\end{eqnarray}

 The most significant point in connection with  (\ref{5.3a})
and (\ref{5.3b}) is that transition from one type of spinor  to
another includes the complex conjugation. This points out that
there  does not  exist analytical  map (in the sense of complex
variable function theory)  which could  be able to relate these
two spinor
 $\xi$   and  $\eta$.  This peculiarity in relation $\xi$ to   $\eta$
is much disguised  when we employ the real-valued variables $U_{a}$ and $V_{a}$.

Let us consider some  additional properties
of the relation  $ V_{a}
= S_{ab} \; U_{b} $:
\begin{eqnarray}
 \left | \begin{array}{c}
V_{4} \\ V_{1} \\ V_{2} \\ V_{3}
\end{array}  \right |
= {1 \over \sqrt{2}}\; \left | \begin{array}{rrrr}
1 & 0 & -1 & 0 \\
0 & 1 &  0 & -1 \\
1 & 0 &  1 & 0 \\
0 & 1 & 0  & 1
\end{array} \right |
\left |  \begin{array}{c}
U_{4} \\ U_{1} \\ U_{2} \\ U_{3}
\end{array}  \right |   .
\label{5.5a}
\end{eqnarray}

It is readily checked that
\begin{eqnarray}
\mbox{det} \; S = +1 \; : \;\;  \Longrightarrow  \;\; S \in
SO(4.R)\; .
\label{5.5b}
\end{eqnarray}

\noindent
Therefore, the matrix  $S$ belongs to the group  $SO(4.R)$.
There exist six elementary rotations in this  group, in the  planes
 $2-3,3-1,1-2$  and  $4-1,3-2,4-3$:
\begin{eqnarray}
S_{2-3}(\phi_{1}) = \left | \begin{array}{cccc}
0 & 0 & 0 & 1\\
0 & 1 & 0 & 0 \\
0 & 0 & \cos \phi_{1} & -\sin \phi_{1}  \\
0 & 0 &  \sin \phi_{1} & \cos \phi_{1}
\end{array} \right |\; , \;\;
S_{3-1}(\phi_{2}) = \left | \begin{array}{cccc}
1 & 0 & 0 & 0\\
0 & \cos \phi_{2} & 0 & -\sin \phi_{2}  \\
0 & 0 & 1 & 0\\
0 &   \sin \phi_{2} & 0&  \cos \phi_{2}
\end{array} \right | \; , \;\;
\nonumber
\end{eqnarray}
\begin{eqnarray}
 S_{1-2}(\phi_{3}) =
\left |
\begin{array}{cccc}
1 & 0 & 0 & 0 \\
0 & \cos \phi_{3} & -\sin \phi_{3} & 0 \\
0 &  \sin \phi_{3} & \cos \phi_{3} & 0\\
0 & 0 & 0 & 1
\end{array} \right | \; , \;\;
S_{4-1}(\beta_{1}) = \left | \begin{array}{cccc}
\cos \beta_{1}  & -\sin \beta_{1} & 0 & 0   \\
 \sin \beta_{1} & \cos \beta_{1} & 0 & 0 \\
0 & 0 & 1 & 0 \\
0 & 0 & 0 & 1
\end{array} \right | \; ,
\nonumber
\end{eqnarray}
\begin{eqnarray}
S_{4-2}(\beta_{2}) = \left | \begin{array}{cccc}
\cos \beta_{2}  & 0 & -\sin \beta_{2}  & 0   \\
0 & 1 & 0 & 0 \\
 \sin \beta_{2} & 0 & \cos \beta_{2} & 0  \\
0 & 0 & 0 & 1
\end{array} \right | \; , \;\;
S_{4-3}(\beta_{3}) = \left | \begin{array}{cccc}
\cos \beta_{3}  & 0 & 0& -\sin \beta_{3}     \\
0 & 1 & 0 & 0 \\
0 & 0 & 1 & 0 \\
 \sin \beta_{3} & 0 & 0 & \cos \beta_{3}   \\
\end{array} \right | \; . \;\;
\label{5.6a}
\end{eqnarray}

Let us  clarify the structure of the  matrix  $S$ in (\ref{5.5a}) in terms of
elementary rotations (\ref{5.6a}):
\begin{eqnarray}
S=
{1 \over \sqrt{2}}\; \left | \begin{array}{rrrr}
1 & 0 & -1 & 0 \\
0 & 1 &  0 & -1 \\
1 & 0 &  1 & 0 \\
0 & 1 & 0  & 1
\end{array} \right |  = {1 \over \sqrt{2}}
 \left | \begin{array}{rrrr}
1 & 0 & -1 & 0 \\
0 & 1 &  0 & 0 \\
1 & 0 &  1 & 0 \\
0 & 0 &  0 & 1
\end{array} \right |
\left | \begin{array}{rrrr}
1 & 0 &  0 & 0 \\
0 & 1 &  0 & -1 \\
0 & 0 &  1 & 0 \\
0 & 1 & 0  & 1
\end{array} \right | \;  .
\label{5.6b}
\end{eqnarray}

\noindent
that is
\begin{eqnarray}
S = S_{4-2}\left({\pi \over 2}\right) \; \; S_{3-1} \left({\pi
\over 4}\right)
=
 S_{4-2}\left({\pi \over 4}\right) \; \; S_{3-1} \left({\pi \over 2}\right) \; .
\label{5.6c}
\end{eqnarray}

Else one  additional test is  possible.
Let us show that the  matrix (\ref{5.5a}), though
being an element of the  group  $SO(4.R)$, nevertheless
it does not belong to its sub-group generated by transformations from  $SU(2)$, acting on spinor
 $\xi$.

Generally speaking, in the  light of eq.  (5.3) the reply to this question is
evident in advance. However it is important in other context. In
 [55-58], instead of the  Kustaanheimo-Stifel variables  $U_{a}$
it was constructed a  complete set of such variables
 $(W_{a},\vec{A})$; more details see in Appendix A.

Transition to the set   $(W_{a},\vec{A})$
is  based on a simple geometrical  idea.
It can be clarified on the familiar case of  spherical coordinates.
Let certain spherical coordinates  be  determined by
\begin{eqnarray}
x= r\sin \theta \sin \phi  ,
y = r \sin \theta \sin\phi ,
 z = r \cos \theta
\nonumber
\end{eqnarray}

\noindent
however it is understandable that you may determine other spherical coordinates
$(r',\theta', \phi')$ related to respective cartesian coordinates  $(x',y',z')$:
\begin{eqnarray}
x' = r'\sin \theta' \sin \phi' ,\;
y' = r' \sin \theta' \sin\phi ', \;
 z' = r' \cos \theta  .
\nonumber
\end{eqnarray}

\noindent
As long as two cartesian sets may be referred to each other through
certain rotation
\begin{eqnarray}
x'_{j} = O_{ji} (c) \; x_{i} \; , \qquad c = (c_{4}, c_{j}) \; ,
\nonumber
\end{eqnarray}

\noindent
spherical sets $(r,\theta,\phi)$ and  $(r',\theta',\phi')$ may be related  as well.
Explicit formulas establishing that connection are going to be not very simple and
interesting. However, the same  idea, being applied to introducing new curvilinear coordinates,
Kustaanheimo-Stifel variables
$(U_{a})$, turns  out to be useful because the formulas relating different sets
$(U_{a})$ and $(U_{a}')$ are quite simple ones.

As shown (see [58] and also Appedix~A,~B of the  present work),
all those new variables $(U_{a}')$  are generated from initial
ones $(U_{a})$ through the matrices of $SU(2)$.

Let us consider action of  $SU(2)$ elements on spatial spinor  $\xi$
in terms of real--valued variables. To this end,
equation
\begin{eqnarray}
\xi' = B(c)\; \xi
\label{5.7a}
\end{eqnarray}

\noindent should be translated to real form. With the help of
\begin{eqnarray}
B(c) =
\left | \begin{array}{rr}
c_{4}  -i c_{3} &  -c_{2}  -i c_{1} \\
c_{2} -i c_{1}  & c_{4}  +i c_{3}
\end{array} \right | ,
\;
\xi = \left | \begin{array}{c}
U_{1} +i\; U_{2} \\
U_{3} + i\; U_{4}
\end{array} \right |  ,
\xi ' = \left | \begin{array}{c}
U_{1}' +i\; U_{2}' \\
U_{3}' + i\; U_{4}'
\end{array} \right |
\label{5.7b}
\end{eqnarray}

\noindent  we get to
\begin{eqnarray}
\left | \begin{array}{c}
U_{4}\; ' \\
U_{1}\;'\\
U_{2}\;'\\
U_{3}\; '
\end{array} \right | =
\left | \begin{array}{rrrr}
c_{4} & -c_{1}  & c_{2} & c_{3} \\
c_{1} & c_{4}  & c_{3} & -c_{2} \\
-c_{2} & -c_{3}  & c_{4} & -c_{1} \\
-c_{3} & c_{2}  & c_{1} & c_{4}
\end{array} \right |
\left | \begin{array}{c}
U_{4} \\
U_{1}\\
U_{2}\\
U_{3}
\end{array} \right |  .
\label{5.7c}
\end{eqnarray}

\noindent
Let us write down the  relation between $\xi (U_{a})$
 $ \eta  (V_{a})$, in the real form too:
\begin{eqnarray}
 \left | \begin{array}{c}
V_{4} \\ V_{1} \\ V_{2} \\ V_{3}
\end{array}  \right |
=
{1 \over \sqrt{2}}\; \left | \begin{array}{cccc}
1 & 0 & -1 & 0 \\
0 & 1 &  0 & -1 \\
1 & 0 &  1 & 0 \\
0 & 1 & 0  & 1
\end{array} \right |
\left |  \begin{array}{c}
U_{4} \\ U_{1} \\ U_{2} \\ U_{3}
\end{array}  \right |   .
\label{5.7d}
\end{eqnarray}

It is seen that the transformation  (\ref{5.7d}) cannot be obtained from
set of matrices (\ref{5.7c}) at any choice of  $(c_{4},c_{j})$.
Therefore, the variables  $U_{a}$  and  $V_{a}$  are essentially different.

\section{Conclusion}

In the work some  relations  between three techniques, Hopf's bundle,
Kustaan\-heimo-Stiefel's
 bundle,  3-space with spinor structure have been examined. The
spinor space is viewed as a real space that is minimally (twice as
much) extended in comparison with an ordinary vector 3-space:  at
this instead of $2\pi$-rotation now only $4\pi$-rotation is taken
to be the  identity transformation in  the geometrical space. With
respect to a given $P$-orientation of an initial unextended
manyfold, vector or pseudovector one, there may be constructed two
different spatial spinors, $\xi$ and $\eta$, respectively. By
definition, those spinors provide us  with points of the extended
space models, each spinor is in the correspondence $2
\longrightarrow 1$ with points of a vector space. For  both models
an explicit parametrization of the spinors  $\xi$ and   $\eta$ by
spherical and parabolic coordinates is given, the parabolic system
turns out to be the  most convenient  for simple defining spacial
spinors. Fours of real-valued coordinates by Kustaanheimo-Stiefel,
$U_{a}$ and $V_{a}$, real and  imaginary parts of complex spinors
$\xi$ and   $\eta$ respectively, obey two quadratic constraints.
So that in both cases, there exists a Hopf's mapping from the part
of 3-sphere  $S_{3}$ into the entire 2-sphere $ S_{2}$. Relation
between two spacial spinor is found: $ \eta =  ( \xi - i \; \sigma
^{2} \xi ^{*} )/\sqrt{2} $, which in terms of Kustaanheimo-Stiefel
variables  $U_{a}$ and  $V_{a}$ is a  linear transformation from
$SO(4.R)$,  which does not enter its sub-group generated by
$SU(2)$-rotation over spinors.

In addition to the main line  of the  present work,
Appendix B  gives some preliminary analysis to the  problem  of finding  $SU(2)$
rotation which connects two different 2-spinors. It is shown that
the use of spinor  description eliminates from the
formalism the  concept of a small group.

\section*{Acknowledgment}
Author is  grateful to all participants of the seminar
 of  Fundamental Interaction Physics Laboratory, and particularly to
Yu.A. Kurochkin and E.A.Tolkachev for interest to work, stimulating  discussions
and advices.


\setcounter{section}{1} \setcounter{equation}{0}

\makeatletter
\renewcommand\theequation{\@Alph\c@section.\@arabic\c@equation}
\makeatother

\section*{Appendix A.
Kustaanheimo-Stifel bundle and covariance
}

 In [58] it was performed generalization\footnote{Below this approach will be
given with some differences in notation.}  for a standard
Kustaanheimo-Stifel formalism:
\begin{eqnarray}
\xi \otimes \xi^{*} = ( r +
x_{j} \sigma^{j})\; ,
\nonumber
\\
x_{j} = {
1\over 2} \; \mbox{sp} \; [\sigma^{j} (\xi \otimes \xi ^{*})] = {1 \over 2} \;
\xi^{+} \sigma^{i} \xi \; ,
\label{A.1b}
\\
 r = {1 \over 2} \; \mbox{sp} \;
(\xi \otimes \xi ^{*}) =
{1 \over  2}  \; \xi^{+} \xi  \; ;
\nonumber
\end{eqnarray}
\begin{eqnarray}
\xi = \left | \begin{array}{c}
U_{1} + i\; U_{2} \\ U_{2} + i \;U_{4}
\end{array} \right | \; ;
\label{A.2}
\end{eqnarray}
\begin{eqnarray}
x_{1} = U_{1} \; U_{3} + U_{2} \;U_{4} \; ,
\nonumber
\\
x_{2} = U_{1}\; U_{4} - U_{2} \;U_{3}\; ,
\nonumber
\\
2\; x_{3} =  U_{1}^{2} + U_{2}^{2} -U_{3}^{2} -U_{4}^{2} \; ,
\nonumber
\\
2\;r=  U_{1}^{2} + U_{2}^{2} +U_{3}^{2} +U_{4}^{2} \;.
\label{A.3}
\end{eqnarray}

\noindent
To a set of  $U_{a}$ there  corresponds a point in the  3-sphere
of the  radius  $\sqrt{2r}$. Significant feature of the  map
 (\ref{A.1b})--(\ref{A.3}) consists in the following:
 all the spinors $\xi$ different in a phase factor  $e^{i\alpha}, \alpha \in [0, 2\pi]$,
will generate the same vector  $\vec{x}$:
\begin{eqnarray}
\xi   \qquad \Longrightarrow  \qquad  \vec{x} \; ,
\qquad
\alpha \in [0, 2\pi]\;, \; \;\; e^{i\alpha}\; \xi ,
\; \Longrightarrow \; \vec{x} \; .
\label{A.4a}
\end{eqnarray}

\noindent
 In terms of  $U_{a}$ this peculiarity looks as
 \begin{eqnarray}
U_{a} \qquad \Longrightarrow \qquad x_{j} \; , \qquad
\nonumber
\\
\left | \begin{array}{c}
\tilde{U}_{1} \\
\tilde{U}_{2}
\end{array} \right | =
\left | \begin{array}{rr}
\cos \alpha & - \sin \alpha \\
+ \sin \alpha & \cos \alpha
\end{array} \right | \left | \begin{array}{c}
U_{1} \\
U_{2}
\end{array} \right | \; ,
\nonumber
\\
\left | \begin{array}{c}
\tilde{U}_{3} \\
\tilde{U}_{4}
\end{array} \right | =
\left | \begin{array}{rr}
\cos \alpha & - \sin \alpha \\
+ \sin \alpha & \cos \alpha
\end{array} \right | \left | \begin{array}{c}
U_{3} \\
U_{4}
\end{array} \right | \; ,
\nonumber
\\
\tilde{U}_{a} \qquad \Longrightarrow \qquad x_{j} \; . \qquad
\label{A.4b}
\end{eqnarray}

\noindent It should be noted, however,  that those $U(1)$-transformations will
take away the points $U_{a}$ from the  surface $U_{1}\;U_{4} + U_{2}\;U_{3}=0$.
Indeed,
\begin{eqnarray}
\tilde{U}_{1}\; \tilde{U}_{4} + \tilde{U}_{2}\; \tilde{U}_{3} =
 \sin 2 \alpha \; ( U_{1}\; U_{3} - U_{2} \; U_{4})  \neq 0  \; .
\label{A.4c}
\end{eqnarray}

\noindent
Only four particular values of $e^{i\alpha}$ leave the  surface $U_{1}U_{4} +
 U_{2}U_{3}=0$ the same:
\begin{eqnarray}
e^{i\alpha}  = +1,-1,+i,-i \; .
\label{A.4d}
\end{eqnarray}

It turned out to  be helpful [35,58] if one  will employ a factorized
form for $U_{a}$:
\begin{eqnarray}
U_{a} = \sqrt{2r}   \; u_{a} ,  \; u_{1}^{2} + u_{2}^{2} + u_{3}^{2} + u_{4}^{2}= +1\; ,
\nonumber
\\
u_{a} = {U_{a} \over \sqrt{2r}} = {U_{a} \over
\sqrt{U_{1}^{2} + U_{2}^{2} +U_{3}^{2} +U_{4}^{2} }} \; .
\label{A.5a}
\end{eqnarray}

\noindent
At this the  spinor $\xi$ from  (\ref{A.2})  becomes
\begin{eqnarray}
\xi =  \sqrt{2r}\;  \left | \begin{array}{c}
u_{1} + i\;u _{2} \\ u_{3} + i \;u_{4}
\end{array} \right | \; ,
\label{A.5b}
\end{eqnarray}

\noindent  and relations    (\ref{A.1b}) and    (\ref{A.3})  give
\begin{eqnarray}
\vec{x} =
{1 \over 2} \;
\xi^{+} \vec{\sigma} \; \xi \;   : \;  \Longrightarrow \;
\vec{n} =
{1 \over \sqrt{2r}} \;
\xi^{+}\; \vec{\sigma} \;  {1 \over \sqrt{2r}} \; \xi \; ,
\nonumber
\end{eqnarray}

\noindent and
\begin{eqnarray}
n_{1} = {x_{1} \over r}  = 2 \;( \; u_{1} \; u_{3} + u_{2} \;u_{4}\;  )\; ,
\nonumber
\\
n_{2} ={x_{2}\over r }  = 2\;(\; u_{1}\; u_{4} - u_{2} \;u_{3}\; )\; ,
\nonumber
\\
n_{3} = {x_{3} \over r}  =  u_{1}^{2} + u_{2}^{2} -u_{3}^{2} -u_{4}^{2} \;,
\nonumber
\\
u_{1}^{2} + u_{2}^{2} + u_{3}^{2} + u_{4}^{2}= +1 \; .
\label{A.5c}
\end{eqnarray}

\noindent
Accounting for the  condition
$u_{a}u_{a}= +1$, with the help of  $u_{a}$  one  can  construct a matrix from
$SU(2)$ group:
\begin{eqnarray}
B(u) = (u_{4} I - i\sigma^{j} u_{j}  ) \in SU(2) \;  .
\nonumber
\end{eqnarray}

\noindent
On direct calculation one readily establishes  [35,58]
the formulas
\begin{eqnarray}
B(u) \sigma^{3} B ^{-1}(u)  =
\left | \begin{array}{cc}
-n_{3} & n_{1} + i n_{2} \\
n_{1} - i n_{2} & +n_{3}  \end{array} \right |
\nonumber
\\
= + n_{1} \sigma^{1} - n_{2} \sigma^{2}
 - n_{3} \sigma^{3} \equiv  \sigma^{1} \; (n_{j} \;\sigma^{j} ) \; \sigma^{1} \; .
\label{A.6a}
\end{eqnarray}

\noindent It is better to rewrite the  formula in the form  (take notice that
$ \pm i\sigma^{1} \in SU(2)$)
\begin{eqnarray}
B(u) \sigma^{3} B^{-1}(u) =
(\pm i\sigma^{1})  ( \vec{n} \; \vec{\sigma} ) (\pm i \sigma^{1} )^{-1}  .
\label{A.6b}
\end{eqnarray}

\noindent
Two  extremes  terms in the right-hand part may be taken to the
left-hand part and also they can be masked by  special notation. Indeed,
\begin{eqnarray}
(\pm i\sigma^{1})^{-1}   B(u)  \;[ (\pm i\sigma^{1}) \sigma^{3}  (\pm i\sigma^{1})^{-1}  ]
 B ^{-1}(u)  (\pm i\sigma^{1})  =- \;
\vec{n}\;   \vec{\sigma}  ; \qquad
\nonumber
\end{eqnarray}

\noindent
from where, with \footnote{
That is  $\hat{u}_{j}$ is given by $u_{j}$ after the  $180^{0}$-rotation over
the axis  $(1,0,0)$).}
\begin{eqnarray}
(\pm i\sigma^{1})^{-1}  B(u)\;   (\pm i\sigma^{1})
= B(u_{4}; u_{1},-u_{2},-u_{3})
\equiv B(\hat{u})\; ,
\label{A.6c}
\end{eqnarray}

\noindent    it follows
\begin{eqnarray}
B(\hat{u}) \sigma^{3} B^{-1}(\hat{u}) = -\; \vec{n} \; \vec{\sigma}\; .
\label{A.7a}
\end{eqnarray}

\noindent This is only another form of the formulas (\ref{A.5c})
determining Kustaanheimo-Stiifel's normalized variables. It is
easily established that the property
  (\ref{A.4a}) of the Kustaanheimo-Stiifel's mapping  after translating to
(\ref{A.7a}) is given by
\begin{eqnarray}
B(\hat{u})  e^{-i\alpha \sigma^{3}}
\sigma^{3}  e^{+i\alpha \sigma^{3}} B^{-1}(\hat{u}) = -\; \vec{n} \; \vec{\sigma}\;
 \; \Longrightarrow
\nonumber
\\
B(\hat{\tilde{u}})\;
\sigma^{3}  \; B^{-1}(\hat{\tilde{u}}) = -\; \vec{n} \; \vec{\sigma}\; . \qquad \qquad
\label{A.7b}
\end{eqnarray}

\noindent
Let us  find how in the form (\ref{A.7a})
will look transformations  of variable $u_{a}$ under  $SU(2)$.
Starting from
\begin{eqnarray}
B(\hat{u}) \sigma^{3} B^{-1}(\hat{u}) = -  \vec{n} \; \vec{\sigma}\;  : \;\;  \Longrightarrow
\nonumber
\\
B(c) B(\hat{u}) \sigma^{3} B^{-1}(\hat{u}) B^{-1}(c)
 = - n_{k}   B(c)  \sigma^{k} B^{-1} (c) \; ,
\nonumber
\end{eqnarray}

\noindent  and with the use of
\begin{eqnarray}
B(c) B(\hat{u}) = B(\hat{u}') \; ,
\label{A.8a}
\end{eqnarray}

\noindent
we get
\begin{eqnarray}
B(\hat{u}')  \sigma^{3}   B^{-1}(\hat{u}')
= -  \sigma^{l} \; [ O_{lk}(c) \;n_{k}  ] =
 -  \vec{\sigma} \; \vec{n}\;'  .
\label{A.8b}
\end{eqnarray}

Taking in mind additional  $\alpha$-rotations (\ref{A.7b}), the formula  (\ref{A.8a})
should  be modified by
\begin{eqnarray}
B(c) B(\hat{u}) \; e^{-i\alpha\sigma^{3}} = B(\hat{u}') \; e^{-i\alpha' \sigma^{3}} \; .
\label{A.8c}
\end{eqnarray}

\noindent
$SU(2)$-transformation of the  matrix (\ref{A.8a}) in the real-valued form will look
\begin{eqnarray}
B(c) B(\hat{u}) = B(\hat{u}')  \qquad \Longrightarrow
\nonumber
\\
\left | \begin{array}{c}
\hat{u}_{4}\; ' \\
\hat{u}_{1}\;'\\
\hat{u}_{2}\;'\\
\hat{u}_{3}\; '
\end{array} \right | =
\left | \begin{array}{rrrr}
c_{4} & -c_{1}  & c_{2} & c_{3} \\
c_{1} & c_{4}  & c_{3} & -c_{2} \\
-c_{2} & -c_{3}  & c_{4} & -c_{1} \\
-c_{3} & c_{2}  & c_{1} & c_{4}
\end{array} \right |
\left | \begin{array}{c}
\hat{u}_{4} \\
\hat{u}_{1}\\
\hat{u}_{2}\\
\hat{u}_{3}
\end{array} \right | \; .
\label{A.8d}
\end{eqnarray}

\noindent There exist useful possibility to modify the  formula
(\ref{A.8d}) as follows:
\begin{eqnarray}
B(\hat{c}) B(u) = B(u') \;: \qquad \Longrightarrow \qquad
\nonumber
\\
\left | \begin{array}{c}
u_{4}\; ' \\
u_{1}\;'\\
u_{2}\;'\\
u_{3}\; '
\end{array} \right | =
\left | \begin{array}{rrrr}
c_{4} & -c_{1}  & -c_{2} & -c_{3} \\
c_{1} & c_{4}  & -c_{3} & c_{2} \\
c_{2} & c_{3}  & c_{4} & -c_{1} \\
c_{3} & -c_{2}  & c_{1} & c_{4}
\end{array} \right |
\left | \begin{array}{c}
u_{4} \\
u_{1}\\
u_{2}\\
u_{3}
\end{array} \right | \; .
\label{A.8e}
\end{eqnarray}

The  formula  (\ref{A.7a}) readily displays remarkable
relationship of Kustaanheimo-Stifel variables with matrices from
orthogonal rotation group $SO(3.R)$. These 3-rotation in the
frames of unitary group is given by
\begin{eqnarray}
O (+c)  = O(-c) =I + 2 \; [\; c_{4} \vec{c}\; ^{\times} + ( \vec{c}\; ^{\times})^{2} \; ]\; ,
\nonumber
\\
 ( \vec{c}\; ^{\times})_{kl}= -\epsilon _{klj} c _{j} \; . \qquad \qquad
\label{A.9a}
\end{eqnarray}

\noindent Because of simple connection that to
3-vector parameter of SO(3.R)
 [60]
\begin{eqnarray}
\vec{C} = {\vec{c} \over c_{4}} \; , \qquad O (+c)  = O(-c)
= I + 2 \; { \; c_{4} \vec{c}\; ^{\times} + ( \vec{c}\; ^{\times})^{2} \over
c_{4}^{2} +  \vec{c}^{2} }
\nonumber
\\
 =
I + 2 \; { \;  \vec{C}\; ^{\times} + ( \vec{C}\; ^{\times})^{2} \over
1  +  \vec{C}^{2} } =O(\vec{C}) \;  ,
\label{A.9b}
\end{eqnarray}

\noindent many advantages of this technique of the use of vector-parameter
[60] preserves for unitary group as well.
This relationships immediately appears as one
takes eq.  (\ref{A.7a}) in the  context of the formula (see (\ref{2.1b}))
\begin{eqnarray}
B^{-1}(c)  \;\sigma^{k} \;B (c) =
O_{kj}(c)  \sigma ^{j} \; , \qquad \Longrightarrow
\nonumber
\\
B(c) \; \sigma^{k} \;B ^{-1}(c) = \sigma ^{l} \; O_{lk}(c)   \; . \qquad
\label{A.10a}
\end{eqnarray}

\noindent If  $k=3$ it takes  on the form
\begin{eqnarray}
   B(c) \; \sigma^{3} \;B ^{-1}(c) = \sigma ^{j} \; O_{j3}(c)   \; .
\label{A.10b}
\end{eqnarray}

\noindent Comparing it with  (\ref{A.7a}),  we get to
\begin{eqnarray}
\sigma ^{j} \; O_{j3}(\hat{u})  = - \sigma^{j} n _{j} : \;
\Longrightarrow \;  n _{j} = -O_{j3}(\hat{u})   \; .
\label{A.10c}
\end{eqnarray}

\noindent
This formula may be checked by direct calculation with the    use of
explicit form of $O_{l3}(\hat{u})$. Starting from
\begin{eqnarray}
O(c) = \left | \begin{array}{ccc}
1 &  0 & 0 \\
0 & 1 & 0 \\
0 & 0 & 1 \end{array} \right | + 2 c_{4}\;
\left | \begin{array}{ccc}
0 &  - c_{3} &  c_{2} \\
c_{3} & 0 &  -c_{1}  \\
-c_{2}  & c_{1}  & 0 \end{array} \right |
+ 2  \left | \begin{array}{ccc}
-c_{2}^{2} -c_{3}^{2} &  c_{1} c_{2} & c_{1}c_{3} \\
c_{2}c_{1} & -c_{3}^{2} -c_{1}^{2} &  c_{2} c_{3} \\
c_{3}c_{1} & c_{3}c_{2} & -c_{2}^{2} -c_{1}^{2}
\end{array} \right | ;
\nonumber
\end{eqnarray}

\noindent or
\begin{eqnarray}
O(\hat{u}) = \left | \begin{array}{ccc}
1 &  0 & 0 \\
0 & 1 & 0 \\
0 & 0 & 1 \end{array} \right | + 2 u_{4}
\left | \begin{array}{ccc}
0 &   u_{3} &  -u_{2} \\
-u_{3} & 0 &  -u_{1}  \\
u_{2}  & u_{1}  & 0 \end{array} \right |
+ 2  \left | \begin{array}{ccc}
-u_{2}^{2} -u_{3}^{2} &  -u_{1} u_{2} & -u_{1}u_{3} \\
-u_{2}u_{1} & -u_{3}^{2} -u_{1}^{2} &  u_{2} u_{3} \\
-u_{3}u_{1} & u_{3}u_{2} & -u_{2}^{2} -u_{1}^{2}
\end{array} \right |  .
\nonumber
\end{eqnarray}

\noindent so that for third column we get
\begin{eqnarray}
\left. \begin{array}{l}
O_{13}(\hat{u}) = -2 (u_{1} u_{3} + u_{2}u_{4}) \; ,\\
O_{23}(\hat{u}) = -2 (u_{1} u_{4} - u_{2}u_{3}) \; ,\\
O_{33}(\hat{u}) = -(u_{1}^{2} +u_{2}^{2} - u_{3}^{2} -u_{4}^{2}) \; .
\end{array} \right.
\label{A.11b}
\end{eqnarray}

\noindent
These relations coincide with  (\ref{A.10c}). These relations  are  manifestly coordinate dependent;
the  task of giving them a covariant form was solved in  [58].
As a base for this eq. (\ref{A.7a}) is taken. It is suggested instead of  (\ref{A.7a})
to use a more general formula
\begin{eqnarray}
B(\hat{w}) \;[\; \vec{A} \; \vec{\sigma}\; ] \; B^{-1}(\hat{w}) = -\;
\vec{n} \; \vec{\sigma}\;  .
\label{A.12}
\end{eqnarray}

Since in (\ref{A.12})  dependence on $A_{j}$ is linear, we may
take the normalization  condition
 $\vec{A}^{\;2} = +1$.
Also it is understandable that
in place of  (\ref{A.12})  more general relation
with 4-component $A^{a}=(A_{4},\vec{A} \; )$ might be taken, however
this does not lead us to any interesting. Indeed
\begin{eqnarray}
B(\hat{w}) \;[ A _{4} I -i \vec{A}  \vec{\sigma}\; ] \; B^{-1}(\hat{w}) =
A_{4} \; I -i (-  \vec{\sigma}  \vec{n}  )   \; .
\nonumber
\end{eqnarray}

The  task is to establish  relation [58]  between
any set  $(w_{a}, \vec{A})$  in  (\ref{A.12}) and the  initial
set $(u_{a})$. To this end, let as start with
\begin{eqnarray}
B(\hat{w}) \; A^{j} \; \sigma^{j}\; B^{-1}(\hat{w}) =
-\; n_{j} \; \sigma^{j}\;  ,
\;\;
A_{1}^{2} + A_{2}^{2} + A_{2}^{2} = +1 \;  .
\label{A.13}
\end{eqnarray}

Let $O(a), a =(a_{4},a_{j})$ be  a  matrix  trans\-forming  the  vector
 $\vec{A}=(A_{1},A_{2},A_{3})$ into the  vector  $\vec{A}_{(+)}=(0,0,+1) $\footnote{
There exists quite elaborated theory of such matrices in the frame of orthogonal
group [59,60], in Section {\bf 8} its spinor  counterpart  will be  considered.}:
\begin{eqnarray}
A^{j}_{(+)} = O_{jl}(a) \; \; A^{l} \; , \;
A^{j} =  O_{jl}^{-1}(a)\; \; A^{l}_{(+)}  \; .
\label{A.14}
\end{eqnarray}

\noindent
With  (\ref{A.14}),   eq. (\ref{A.13})  becomes
\begin{eqnarray}
B(\hat{w}) \; \sigma^{j}\;  O_{jl}^{-1}(a) \; A^{l}_{(+)}   \;  B^{-1}(\hat{w}) =
-\; n_{j} \; \sigma^{j} \;  .
\nonumber
\end{eqnarray}

\noindent
From this, with the use of
\begin{eqnarray}
\sigma^{j} \;O_{jl}^{-1}(a)     = B^{-1}(a) \; \sigma^{l} \; B(a) \; ,
\nonumber
\end{eqnarray}

\noindent one  gets
\begin{eqnarray}
B(\hat{w}) \; B^{-1}(a) \; \sigma^{l} \;   A^{l}_{(+)}  \;B(a)    \;  B^{-1}(\hat{w}) =
-\; \vec{n}\; \vec{\sigma}\;  ,
\nonumber
\end{eqnarray}

\noindent  or
\begin{eqnarray}
[ B(\hat{w})  B(\bar{a}) ] \; \sigma^{3} \; [B(\hat{w}) \; B(\bar{a}) ]^{-1} = -
\vec{n}\; \vec{\sigma}\;  .
\label{A.15}
\end{eqnarray}

\noindent Comparing this  with  (\ref{A.7a}):
\begin{eqnarray}
B(\hat{u}) \sigma^{3} B^{-1}(\hat{u}) = -\; \vec{n} \; \vec{\sigma}\; ,
\nonumber
\end{eqnarray}

\noindent we arrive at condition connecting  $(u)$ and $(w,\vec{A})$:
\begin{eqnarray}
B(\hat{u}) = B(\hat{w}) \; B^{-1}(a) , \quad
B(\hat{w}) =  B(\hat{u}) \; B(a)  \; ,
\label{A.16a}
\end{eqnarray}

\noindent or  a more precise form
\begin{eqnarray}
B(\hat{u}) e^{-i\alpha \sigma^{3}}  = B(\hat{w}) \; B^{-1}(a)
e^{-i \beta \sigma^{3}} \;\;  \Longrightarrow
\nonumber
\\
B(\hat{w}) =  B(\hat{u})  e^{-i(\alpha -\beta) \sigma^{3}}\; B(a)  \; . \qquad
\label{A.16b}
\end{eqnarray}

\noindent
Mention that   $a=(a_{4}, a_{j})$ is defined  by
\begin{eqnarray}
A^{j}_{(+)} = O_{jl}(a) \; A^{l}\; , \; \vec{A} = (A_{1},A_{2},A_{3}) \; ,
\nonumber
\\
A_{1}^{2} + A_{2}^{2} + A_{2}^{2} = +1 , \; \vec{A}_{0}=(0,0,+1) \; .
\label{A.16c}
\end{eqnarray}

Let us write down the relationship between
  $(u),(w)$-variables and symmetry transformation over $(u)$:
\begin{eqnarray}
 B(\hat{w}) =  B(\hat{u}) \; B(a) \; ,
\label{A.17a}
\\
 B(\hat{u}') =B(c) B(\hat{u})  \; .
\label{A.17b}
\end{eqnarray}

\noindent One  may identify м $(\hat{w}) = (\hat{u}')$,
so that a symmetry  operation that generates   $(w)$ from  $(u)$ is found:
\begin{eqnarray}
B(c) = B(\hat{u}) B(a) B^{-1}(\hat{u}) \; ,
\label{A.17c}
\end{eqnarray}

\noindent  or its more  precise form
\begin{eqnarray}
B(c) = B(\hat{u})  e^{-(\alpha - \beta)\sigma^{3}} B(a)
[  B(\hat{u}) e^{-(\alpha - \beta)\sigma^{3}} ]^{-1}  .
\nonumber
\end{eqnarray}

For more understanding properties of   $(w,\vec{A})$-variables it is useful to perform some  additional
manipulation with formulas. The  idea is to change the left-right side  of
(\ref{A.12}) to the form of  (\ref{A.7a}), at performing this we may expect appearance
in the right-hand side a term with a new vector
 $\vec{n}\;'$ that is specially rotated with respect to the  initial   $\vec{n}$.

To this end, let us express from (\ref{A.12})  the combination
 $ \; \vec{A} \; \vec{\sigma}\;   $:
\begin{eqnarray}
 A^{j}  \; \sigma^{j}  = - \;  n_{j}  \; \;B^{-1} (\hat{w}) \; \sigma^{j}   \;
B(\hat{w}) \;.
\label{A.18a}
\end{eqnarray}

\noindent
From where, with the use of
\begin{eqnarray}
B^{-1} (\hat{w}) \; \sigma^{j}   \;
B(\hat{w})
= O _{jl}(\hat{w}) \; \sigma^{l} \; ,
\nonumber
\end{eqnarray}

\noindent we get
\begin{eqnarray}
 A^{j}  \; \sigma^{j}  =  -\; n _{j} \;
 \; O _{jl}(\hat{w}) \; \sigma^{l} \; .
\label{A.18b}
\end{eqnarray}

Now let us carry out a special transformation giving
on the left the  term  $\sigma^{3} $. For this, let us multiply  eq.(\ref{A.18b}) by
$B(a)$ from the left and by   $B^{-1}(a)$  from the right:
\begin{eqnarray}
A^{j}    B(a) \sigma^{j}  B^{-1} (a)  =  - n_{j}
O _{jl}(\hat{w})    B(a) \sigma^{l} B^{-1} (a) \; ,
\nonumber
\end{eqnarray}

\noindent
and further
\begin{eqnarray}
\sigma^{k} \; [ \; O(a)_{kj}\;A^{j}\; ] =
-\; n_{j} \;
 \; O _{jl}(\hat{w}) \;  O^{-1}_{lk}(a) \; \sigma^{k} \; .
\nonumber
\end{eqnarray}

\noindent
Since
\begin{eqnarray}
[ \;  O(a)_{kj}\;A^{j}  \;  ]  = (0,0,+1) \; ,
\nonumber
\end{eqnarray}

\noindent the previous relation will take the form
\begin{eqnarray}
\sigma^{3} =
-\; n_{j} \;
 \; O _{jl}(\hat{w}) \;  O^{-1}_{lk}(a) \; \sigma^{k} \; .
\label{A.19b}
\end{eqnarray}

\noindent
Now,  let us multiply eq.  (\ref{A.19b}) by  $B(\hat{w})$ from the left and  by
 $B^{-1}(\hat{w})$  from the right:
\begin{eqnarray}
B(\hat{w}) \sigma^{3} B^{-1}(\hat{w})
= -
n_{j}
\; O _{jl}(\hat{w})   O^{-1}_{lk}(a)   B(\hat{w}) \sigma^{k}B^{-1}(\hat{w} )  \; .
\label{A.20a}
\end{eqnarray}

\noindent
from where it follows
\begin{eqnarray}
B(\hat{w}) \sigma^{3} B^{-1}(\hat{w})
=
-\; n_{j} \;
 \; O _{jl}(\hat{w}) \;  O^{-1}_{lk}(a) \;  O^{-1}_{ki} (\hat{w}) \sigma^{i}   \; ,
\label{A.20b}
\end{eqnarray}

\noindent
or
\begin{eqnarray}
B(\hat{w}) \sigma^{3} B^{-1}(\hat{w})
=  - \sigma^{j}  [
 O_{ik}(\hat{w}) O_{kl}(a) O^{-1}_{lj} (\hat{w})  ] n_{j} .
\label{A.20c}
\end{eqnarray}

\noindent
In index-free form it looks as
\begin{eqnarray}
B(\hat{w}) \sigma^{3} B^{-1}(\hat{w}) = -  \vec{\sigma}  [
O(\hat{w}) O(a) O^{-1} (\hat{w})  ] \vec{n}  .
\label{A.21}
\end{eqnarray}

\noindent
Relation  (\ref{A.21}), with the  use of designation
\begin{eqnarray}
\vec{n}\;'  \equiv
\; [\;
 O(\hat{w}) O(a) O^{-1} (\hat{w}) \; ]\; \vec{n} \;  ,
\label{A.22}
\end{eqnarray}

\noindent will take the form required:
\begin{eqnarray}
 B(\hat{w}) \sigma^{3} B^{-1}(\hat{w}) = - \;  \vec{\sigma} \;
 \;\vec{n}\; ' \; .
\label{A.23}
\end{eqnarray}

It is evident that both definitions of
$(w)$ -- on the  base of    (\ref{A.23}) or
(\ref{A.12}):
\begin{eqnarray}
B(\hat{w}) \;[\; \vec{A} \; \vec{\sigma}\; ] \; B^{-1}(\hat{w}) = -\;
\vec{n} \; \vec{\sigma}\;  .
\label{A.24}
\end{eqnarray}

\noindent are  absolutely.
equivalent. However, the form  (\ref{A.23}) distinctly reveals geometrical meaning of
existence a great many of Kustaanheimo-Stifel's coordinates in place of a single one.

\setcounter{section}{2} \setcounter{equation}{0}

\section*{Appendix B. \\On unitary gauges of   spinors }

\hspace{5mm} Normalized  pseudo vector
$\vec{n}, \; \vec{n}\; ^{2} = +1$ can be formed up  from a normalized spinor $\Psi$
in accordance with
\begin{eqnarray}
\Psi  \otimes \Psi^{*} = {1 \over 2} (1
+ \sigma^{n} \; \vec{n} ) \; ,
\label{B.1a}
\nonumber
\\
 1 =  \mbox{sp} \;
(\Psi \otimes \Psi ^{*}) =  \Psi^{+} \Psi = \mbox{inv} \; ,
\nonumber
\\
n_{j} =  \mbox{sp} \; [\sigma^{j} (\Psi \otimes \Psi ^{*})] = \Psi^{+} \sigma^{i} \;\Psi \; .
\nonumber
\end{eqnarray}

The spinor $\Psi$ corresponding to  $\vec{n}$ is given by
\begin{eqnarray}
\Psi = {1 \over \sqrt{2}} \;
\left | \begin{array}{c}
\sqrt{1 +n_{3}} e^{-i\gamma/2} \\[2mm]
\sqrt{1 -n_{3}} e^{+i\gamma/2}
\end{array} \right |  ,
\nonumber
\\
 e^{i\gamma} = {n_{1} + in_{2} \over \sqrt{n_{1}^{2} +
n_{2}^{2}}} \; , \;
\gamma \in [ \; -2 \pi, \; +2\pi \; ]\; .
\label{B.1b}
\end{eqnarray}

At two points of the sphere,  $(0,0,+1)$ and  $(0,0,-1)$,  the spinor
has peculiarity  -- it is not single-valued function of points on the sphere:
\begin{eqnarray}
\Psi^{(+)}  =
\left | \begin{array}{c}
e^{-i\Gamma/2} \\[2mm]
0
\end{array} \right |  , \;
\Psi^{(-)}  =
\left | \begin{array}{c}
0 \\[2mm]
e^{+i\Gamma/2}
\end{array} \right | \; ,
\label{B.1c}
\end{eqnarray}

By definition, $(+)$-unitary gauge of a vector $\vec{n}$ is result of
rotating the vector to the  positive  axis  $z$:
\begin{eqnarray}
\vec{n} = (n_{1},n_{2},n_{3}) \; \stackrel{O(c)}{\Longrightarrow} \;
\vec{n}_{(+)} = (0,0,+1)\; .
\label{B.2a}
\end{eqnarray}

By definition, $(-)$-unitary gauge of a vector $\vec{n}$ is result of
rotating the vector to the  negative axis  $z$:
\begin{eqnarray}
\vec{n} = (n_{1},n_{2},n_{3}) \; \stackrel{O(c)}{\Longrightarrow} \;
\vec{n}_{(-)} = (0,0,-1)\; .
\label{B.2b}
\end{eqnarray}

\vspace{5mm}

There may be defined a counterpart of this concept in spinor
formalism:

$(+)$-unitary gauge of a spinor $\Psi$  is result of $SU(2)$-rotating the spinor
$\Psi$ to the form
\begin{eqnarray}
\Psi = {1 \over \sqrt{2}} \;
\left | \begin{array}{c}
\sqrt{1 +n_{3}}\; e^{-i\gamma/2} \\[2mm]
\sqrt{1 -n_{3}}\; e^{+i\gamma/2}
\end{array} \right | \qquad  \stackrel{B(c)}{\Longrightarrow} \qquad
\Psi^{(+)}  =
\left | \begin{array}{c}
e^{-i\Gamma/2} \\[2mm]
0
\end{array} \right | \; ; \qquad \qquad
\label{B.3a}
\end{eqnarray}

$(-)$-unitary gauge of a spinor $\Psi$  is result of $SU(2)$-rotating the spinor
$\Psi$ to the form
\begin{eqnarray}
\Psi = {1 \over \sqrt{2}} \;
\left | \begin{array}{c}
\sqrt{1 +n_{3}}\; e^{-i\gamma/2} \\[2mm]
\sqrt{1 -n_{3}}\; e^{+i\gamma/2}
\end{array} \right | \; \qquad \stackrel{B(c)}{\Longrightarrow}
\qquad
\Psi^{(-)}  =
\left | \begin{array}{c}
0\\[2mm]
e^{+i\Gamma/2}
\end{array} \right | \; . \qquad \qquad
\label{B.3b}
\end{eqnarray}

First, let us consider  the case of
$(+)$-unitary gauge of spinor; from which
its vector form cam be readily produced.
The main equation to solve is
\begin{eqnarray}
B(c) \Psi = \Psi^{(+)} \; .
\label{B.4a}
\end{eqnarray}

\noindent In real-valued Kustaanheimo-Stifel repre\-sen\-tation
this  equation takes on the form
\begin{eqnarray}
\left | \begin{array}{rrrr}
c_{4} & -c_{1}  & c_{2} & c_{3} \\
c_{1} & c_{4}  & c_{3} & -c_{2} \\
-c_{2} & -c_{3}  & c_{4} & -c_{1} \\
-c_{3} & c_{2}  & c_{1} & c_{4}
\end{array} \right |
\left | \begin{array}{c}
u_{4} \\
u_{1}\\
u_{2} \\
u_{3}
\end{array} \right |  =
\left | \begin{array}{c}
0 \\
+ \cos (\Gamma /2) \\
- \sin (\Gamma /2) \\
0
\end{array} \right |  \; .
\nonumber
\end{eqnarray}
\begin{eqnarray}
\label{B.4b}
\end{eqnarray}

The  problem is to find  $c_{a}$ at given  $u_{a}$ and $e^{i\Gamma/2}$.
Equation  (\ref{B.4b}) can be re-written as
\begin{eqnarray}
\left | \begin{array}{rrrr}
c_{4} & -c_{1}  & c_{2} & c_{3} \\
c_{1} & c_{4}  & c_{3} & -c_{2} \\
-c_{2} & -c_{3}  & c_{4} & -c_{1} \\
-c_{3} & c_{2}  & c_{1} & c_{4}
\end{array} \right |
\left | \begin{array}{c}
u_{4} \\
u_{1}\\
u_{2} \\
u_{3}
\end{array} \right |
 =
\left | \begin{array}{cccc}
\cos {\Gamma \over 2}  & 0  & 0 & \sin {\Gamma \over 2}  \\
0  &  \cos {\Gamma \over 2}    & \sin {\Gamma \over 2}  & 0 \\
0 & -\sin {\Gamma \over 2}    & \cos {\Gamma \over 2}  & 0 \\
-\sin {\Gamma \over 2}  & 0  & 0 & \cos {\Gamma \over 2}
\end{array} \right |
\left | \begin{array}{c}
0 \\
1 \\
0 \\
0
\end{array} \right |  \; ;
\label{B.5a}
\end{eqnarray}

\noindent or in matrix form
\begin{eqnarray}
S(c) \; u = S\left( \cos {\Gamma  \over 2},0,0, \sin {\Gamma  \over 2}\right)
\; u^{(+)}_{0} \; ,
\nonumber
\\
B(c) \Psi = B \left( \cos {\Gamma  \over 2} ,0,0, \sin {\Gamma  \over 2}
\right) \; \Psi^{(+)}_{0} \;.
\label{B.5b}
\end{eqnarray}

\noindent From here it  follows
\begin{eqnarray}
S\left( \cos {\Gamma  \over 2},0,0, - \sin {\Gamma  \over 2} \right) \;  S(c) \; u =  u^{(+)}_{0} \; ,
\nonumber
\\
B\left( \cos {\Gamma  \over 2} ,0,0, - \sin {\Gamma  \over 2} \right)\;  B(c) \; \Psi =  \psi^{(+)}_{0} \;
\label{B.5c}
\end{eqnarray}

\noindent or
\begin{eqnarray}
S(a) \; u =  u^{(+)}_{0} \; ,\;\;
B(a) \; \Psi  =  \Psi^{(+)}_{0} \; .
\label{B.6a}
\end{eqnarray}

\noindent  The designation is used:
\begin{eqnarray}
S(a) = S\left(\cos {\Gamma \over 2},0,0, - \sin {\Gamma \over 2}\right) \; S(c)\; ,
\nonumber
\\
B(a) = B\left(\cos {\Gamma \over 2},0,0, - \sin {\Gamma \over 2} \right) \; B(c)\; ;
\label{B.6b}
\end{eqnarray}

\noindent  or in explicit form
\begin{eqnarray}
a_{1} = \; \cos {\Gamma \over 2} \; c_{1} + \sin {\Gamma \over 2}\; c_{2} \;  ,
\qquad
\;
a_{2} = -\sin {\Gamma \over 2} \; c_{1} + \cos {\Gamma \over 2}  \;c_{2} \; ,
\nonumber
\\
a_{3} = \; \cos {\Gamma \over 2} \; c_{3} - \sin {\Gamma \over 2}\; c_{4} \; ,
\qquad
\;
a_{4} = \; \sin {\Gamma \over 2}  \;c_{3} + \cos {\Gamma \over 2} \;c_{4} \; .
\label{B.6c}
\end{eqnarray}

Equation (\ref{B.6a}) gives
\begin{eqnarray}
\left | \begin{array}{rrrr}
a_{4} & -a_{1}  & a_{2} & a_{3} \\
a_{1} & a_{4}  & a_{3} & -a_{2} \\
-a_{2} & -a_{3}  & a_{4} & -a_{1} \\
-a_{3} & a_{2}  & a_{1} & a_{4}
\end{array} \right |
\left | \begin{array}{c}
u_{4} \\
u_{1}\\
u_{2} \\
u_{3}
\end{array} \right |  =
\left | \begin{array}{c}
0 \\
1 \\
0 \\
0
\end{array} \right |  .
\label{B.7a}
\end{eqnarray}

\noindent
After ordering it with respect to unknown $(a_{4},\vec{a})$, it  looks as a linear system
\begin{eqnarray}
\left | \begin{array}{rrrr}
u_{4} & -u_{1}  & u_{2} & u_{3} \\
u_{1} & u_{4}  & -u_{3} & u_{2} \\
u_{2} & -u_{3}  & -u_{4} & -u_{1} \\
u_{3} & u_{2}  & u_{1} & -u_{4}
\end{array} \right |
\left | \begin{array}{c}
a_{4} \\
a_{1}\\
a_{2} \\
a_{3}
\end{array} \right |  =
\left | \begin{array}{c}
0 \\
1 \\
0 \\
0
\end{array} \right |  .
\label{B.7b}
\end{eqnarray}

\noindent
Its main determinant equals $ +1$:
\begin{eqnarray}
\Delta = (u_{1}^{2} + u_{2}^{2} + u_{3}^{2} + u_{4}^{2})^{2} = +1 \; ;
\nonumber
\end{eqnarray}

\noindent therefore a single solution exists:
\begin{eqnarray}
a_{4} = + u_{1} \; , \;\; a_{1} = + u_{4} \; ,
\nonumber
\\
a_{2} = - u_{3} \; , \;\; a_{3} = + u_{2} \; ,
\label{B.7c}
\end{eqnarray}

\noindent
which is readily checked by  (see (\ref{B.7a}))
\begin{eqnarray}
\left | \begin{array}{rrrr}
u_{1} & -u_{4}  & -u_{3} & u_{2} \\
u_{4} & u_{1}  & u_{2} & u_{3} \\
u_{3} & -u_{2}  & u_{1} & -u_{4} \\
-u_{2} & -u_{3}  & u_{4} & u_{1}
\end{array} \right |
\left | \begin{array}{c}
u_{4} \\
u_{1}\\
u_{2} \\
u_{3}
\end{array} \right |  =
\left | \begin{array}{c}
0 \\
1 \\
0 \\
0
\end{array} \right |  \; .
\label{B.8a}
\end{eqnarray}

\noindent
The  solution obtained after translating to spinor form becomes
\begin{eqnarray}
 B(a) =
u_{1} -i\sigma^{1} u_{4} +
i\sigma^{1} u_{3} - i\sigma^{3} u_{2}  , \;
B(a) \Psi = \Psi^{+}_{0} \; ,
\nonumber
\\
\left | \begin{array}{cc}
u_{1} -iu_{2} & u_{3} -i u_{4} \\
-u_{3} -i u_{4} & u_{1} +iu_{2}
\end{array} \right |
\left | \begin{array}{c}
u_{1} + iu_{2} \\u_{3} + i u_{4}
\end{array} \right | =
\left | \begin{array}{c}
1 \\ 0
\end{array} \right |  ,
\nonumber
\end{eqnarray}

\noindent Thus, the problem of  $(+)$-unitary gauge of spinors is solved
 (with eq.~(\ref{B.5c})):
\begin{eqnarray}
B(a) \Psi = \Psi^{(+)}_{0}  , \qquad
B(a)  = B\left(  \cos {\Gamma \over 2},0,0, - \sin {\Gamma \over 2} \right)  B(c) \; ,
\nonumber
\\
 B(c)  \Psi = \Psi^{+}  \; , \qquad
B(c) = B\left( \cos {\Gamma \over 2},0,0,  \sin {\Gamma \over 2} \right)  \; B(a)  .
\label{B.9a}
\end{eqnarray}

\noindent
Explicit form of $(c)$ is
\begin{eqnarray}
c_{1} = \; \cos {\Gamma \over 2} \; u_{4} + \sin {\Gamma \over 2}\; u_{3}
,\qquad
c_{2} = \;\sin {\Gamma \over 2} \; u_{4} - \cos {\Gamma \over 2}\; u_{3}
 ,
\nonumber
\\
c_{3} = \;\cos {\Gamma \over 2} \; u_{2} + \sin {\Gamma \over 2}\; u_{1}
  ,\qquad
  c_{4} = -\sin {\Gamma \over 2} \; u_{2} + \cos {\Gamma \over 2}\; u_{1}
\label{B.9b}
\end{eqnarray}

\noindent
and  $B(c)$ looks
\begin{eqnarray}
B(c) = \left | \begin{array}{rr}
e^{-i\Gamma /2}  (u_{1} -i u_{2})  &  e^{-i\Gamma /2}  (u_{3} -i u_{4}) \\
-e^{+i\Gamma /2}  (u_{1} +i u_{2}) & e^{+i\Gamma /2}  (u_{1} +i u_{2})
\end{array} \right |
\nonumber
\end{eqnarray}

\noindent which can be  readily verified by
\begin{eqnarray}
\left | \begin{array}{rr}
e^{-i\Gamma /2} \; (u_{1} -i\; u_{2})  &  e^{-i\Gamma /2} \; (u_{3} -i\; u_{4}) \\
-e^{+i\Gamma /2} \; (u_{1} +i\; u_{2}) & e^{+i\Gamma /2} \; (u_{1} +i\; u_{2})
\end{array} \right |
\left | \begin{array}{c}
u_{1} + i\;u_{2} \\u_{3} + i\; u_{4}
\end{array} \right | =
\left | \begin{array}{c}
e^{-i\Gamma/2}  \\ 0
\end{array} \right |  . \qquad
\nonumber
\end{eqnarray}

It is  simple thing to show how this theory of unitary spinor
gauge is referred to the well-known theory of flat orthogonal
rotations translating the  vector $\vec{n}=(n_{1}, n_{2},n_{3})$
into $\vec{n}_{(+)}=(0,0,+1)$ (see  [59,60]). At the  first place
it should be  noted that presence of a  phase  factor $e^{-i
\Gamma/2}$ at  $\Psi^{+}$ by no means influences the  vector
associated with this  spinor:
\begin{eqnarray}
\Psi^{+} \qquad \Longrightarrow \qquad  \vec{n}_{(+)} =(0,0,+1) \; ,
\nonumber
\\
\Psi^{+}_{0}  \qquad \Longrightarrow \qquad  \vec{n}_{(+)} =(0,0,+1) \; .
\label{B.11a}
\end{eqnarray}

\noindent This  circumstance  may be  used as follows: let this phase be such that
 $c_{3}$ vanishes:
 \begin{eqnarray}
c_{3}=0  \; \Longrightarrow \; \cos {\Gamma \over 2 }  u_{2} + \sin {\Gamma \over 2}
u_{1} =0 ,
\label{B.11b}
\end{eqnarray}

\noindent that is
\begin{eqnarray}
\tan {\Gamma \over 2} = -\; {u_{2} \over u_{1} } \; , \;\;\; {\Gamma \over 2} \in [-\pi, + \pi ]\; .
\nonumber
\end{eqnarray}

\noindent
Such a special choice of the  phase  does not contain any  puzzle. Indeed,
out of the  axis  $z$  for real-valued   $u_{1}$ and   $u_{2}$  we have expressions
\begin{eqnarray}
u_{1}  = \sqrt{1-n_{3}}\;
\cos {\gamma \over 2}  \; , \qquad
 u_{2} = -\; \sqrt{1-n_{3}}\;
 \sin { \gamma \over 2 }  \; .
\label{B.11c}
\end{eqnarray}

\noindent Now preserving  $\gamma$ constant one  can change
 $n_{3} \rightarrow +1$, so that
the variable  $\gamma$ turns  out to be a  "mute" \hspace{2mm}  coordinate $\Gamma$:
\begin{eqnarray}
\Gamma = \gamma\; .
\label{B.11d}
\end{eqnarray}

\noindent
In other word, the choice  (\ref{B.11b})  is simple acceptance of equations
 (\ref{B.11c})--(\ref{B.11d}).

With  (\ref{B.11b}),  expression for a spinor matrix   $B(c)$  becomes simpler,
and  what is  more, its  vector counterpart, orthogonal matrix $O(c)$ is much  simplified
as well. Indeed,
\begin{eqnarray}
c_{3}= 0 \; , \qquad  c_{4} = \cos ( \Gamma  /2)   { u_{1}^{2}  + u_{2}^{2} \over u_{1} } \; ,
\nonumber
\\
c_{1} = \cos ( \Gamma /  2)  {u_{4} u_{1} -u_{2} u_{3} \over u_{1} } \; ,
\;\;
c_{2} = -\cos ( \Gamma / 2)   { u_{3} u_{1} + u_{2} u_{3} \over u_{1} } \; ,
\label{B.12a}
\end{eqnarray}

\noindent 3-vector parameter becomes
\begin{eqnarray}
C_{1} = {c_{1} \over c_{4} } = + \;{u_{4} u_{1} -u_{2} u_{3} \over u_{1}^{2} + u_{2}^{2}  } =
+\; {n_{2} \over 1+ n_{3}} \; ,
\nonumber
\\
C_{2} = {c_{2} \over c_{4} } =  - {u_{1} u_{3} +u_{2} u_{4} \over u_{1}^{2} + u_{2}^{2}  }
=
-\; {n_{1} \over 1+ n_{3}} \; ,
\;\;
 C_{3} =0 \; .
\label{B.12b}
\end{eqnarray}

The  obtained  $C_{i}(n)$ is  the  well-known expression for
3-vector parameter of a flat rotation  from  $SO(3.R)$, translating vector
 $\vec{n}$ onto positive axis  $z$.
This  $\vec{C}$ is directed along
\begin{eqnarray}
\vec{n} \times \vec{n}_{(+)} =
\left | \begin{array}{ccc}
\vec{e}_{1} & \vec{e}_{2} & \vec{e}_{3} \\
n_{1} & n_{2} & n_{3} \\
0 & 0 & +1
\end{array} \right | = n_{2} \; \vec{e}_{1} - n_{1}  \;\vec{e}_{2}  .
\nonumber
\end{eqnarray}

\noindent
With the help of
\begin{eqnarray}
n_{1} = \sin \theta \cos \phi  ,\;
n_{2}  = \sin \theta \; \sin \phi  , \;
n_{3} = \cos \theta \; ,
\nonumber
\end{eqnarray}

\noindent the modulus of the vector is  found immediately:
\begin{eqnarray}
\mid \vec{C} \mid =   \sqrt{{n_{1}^{2} +n_{2}^{2} \over (1+n_{3})^{2}}} = \sqrt{{
\sin^{2} \theta \over (1 + \cos \theta)^{2}}} = \tan {\theta \over 2}  .
\nonumber
\end{eqnarray}

\vspace{5mm}
In the same  line,  one  can consider  $(-)$-unitary spinor and vector gauges;
at this the most calculation need not do be repeated. The  principal formulas are:
\begin{eqnarray}
\Psi = \left | \begin{array}{c}
u_{1} + i\; u_{2} \\
u_{3} + i\; u_{4}
\end{array} \right |  ,  \qquad
\Psi^{(-)} = \left | \begin{array}{c}
0 \\
e^{+i\Gamma /2} \end{array} \right |  , \;
\Psi^{(-)}_{0} = \left | \begin{array}{c}
0 \\
1 \end{array} \right |  ,
\nonumber
\\
B(c) \Psi = \Psi^{(-)} \; , \qquad
\left | \begin{array}{rrrr}
c_{4} & -c_{1}  & c_{2} & c_{3} \\
c_{1} & c_{4}  & c_{3} & -c_{2} \\
-c_{2} & -c_{3}  & c_{4} & -c_{1} \\
-c_{3} & c_{2}  & c_{1} & c_{4}
\end{array} \right |
\left | \begin{array}{c}
u_{4} \\
u_{1}\\
u_{2} \\
u_{3}
\end{array} \right |
=
\left | \begin{array}{c}
\sin (\Gamma /2)  \\
0 \\
0 \\
 \cos (\Gamma /2)
\end{array} \right |   ,
\;\;
\nonumber
\\
B \left(\cos {\Gamma \over 2}   , 0, 0, +\;  \sin {\Gamma \over 2}\right) \;\Psi^{(-)}_{0}=
 \Psi^{(-)}\; , \;\;
\nonumber
\\
B \left( \cos {\Gamma \over 2}  , 0, 0, - \;\sin {\Gamma \over 2}\right) \; B(c)\; \Psi =
\Psi^{(-)}_{0} ,
\; \; B(a)\; \Psi = \Psi^{(-)}_{0}. \label{B.14a}
\end{eqnarray}

\noindent
Equations from  (\ref{B.14a}) after translating to real-valued form will look
\begin{eqnarray}
\left | \begin{array}{rrrr}
a_{4} & -a_{1}  & a_{2} & a_{3} \\
a_{1} & a_{4}  & a_{3} & -a_{2} \\
-a_{2} & -a_{3}  & a_{4} & -a_{1} \\
-a_{3} & a_{2}  & a_{1} & a_{4}
\end{array} \right |
\left | \begin{array}{c}
u_{4} \\
u_{1}\\
u_{2} \\
u_{3}
\end{array} \right |  =
\left | \begin{array}{c}
0 \\
0 \\
0 \\
1
\end{array} \right |  ;
\nonumber
\end{eqnarray}

\noindent
which after re-grouping it  with respect to variables  $(a_{4},\vec{a})$  looks
\begin{eqnarray}
\left | \begin{array}{rrrr}
u_{4} & -u_{1}  & u_{2} & u_{3} \\
u_{1} & u_{4}  & -u_{3} & u_{2} \\
u_{2} & -u_{3}  & -u_{4} & -u_{1} \\
u_{3} & u_{2}  & u_{1} & -u_{4}
\end{array} \right |
\left | \begin{array}{c}
a_{4} \\
a_{1}\\
a_{2} \\
a_{3}
\end{array} \right |  =
\left | \begin{array}{c}
0 \\
0 \\
0 \\
1
\end{array} \right |  \; ;
\label{B.14b}
\end{eqnarray}
\noindent  its solution
\begin{eqnarray}
a_{4} = + u_{3} \; , \;\; a_{1} = + u_{2} \; ,\;
a_{2} = + u_{1} \; , \;\; a_{3} = - u_{4} \; .
\label{B.14c}
\end{eqnarray}

\noindent
Corresponding  $(c_{4}, \vec{c})$:
\begin{eqnarray}
c_{1} = \cos {\Gamma \over 2} \; u_{2} - \sin {\Gamma \over 2}\; u_{1}
\; ,\qquad
c_{2} = \sin {\Gamma \over 2} \; u_{2} + \cos {\Gamma \over 2}\; u_{1}
 \; ,
\nonumber
\\
c_{3} =
-\cos {\Gamma \over 2} \; u_{4} + \sin {\Gamma \over 2}\; u_{3}
 \; ,
\qquad
c_{4} = \sin {\Gamma \over 2} \; u_{4} + \cos {\Gamma \over 2}\; u_{3}\;
\label{B.15a}
\end{eqnarray}

\noindent
 spinor  matrix  $B(c)$:
\begin{eqnarray}
B = \left | \begin{array}{rr}
e^{-i\Gamma /2}  (u_{3} +i u_{4})  &  -e^{-i\Gamma /2}  (u_{1} +i u_{2}) \\
e^{+i\Gamma /2}  (u_{1} -i u_{2}) & e^{+i\Gamma /2} (u_{3} -i u_{4})
\end{array} \right |  ;
\nonumber
\end{eqnarray}

\noindent  gives equation $B(c) \Psi = \Psi^{(-)}$ which is  checked by
\begin{eqnarray}
\left | \begin{array}{rr}
e^{-i\Gamma /2} \; (u_{3} +i\; u_{4})  &  -e^{-i\Gamma /2} \; (u_{1} +i\; u_{2}) \\
e^{+i\Gamma /2} \; (u_{1} -i\; u_{2}) & e^{+i\Gamma /2} \; (u_{3} -i\; u_{4})
\end{array} \right |
 \left | \begin{array}{c}
u_{1} + i\; u_{2} \\
u_{3} + i\; u_{4}
\end{array} \right | =
\left | \begin{array}{c}
0 \\ e^{+i\Gamma/2} \end{array} \right |  . \qquad
\nonumber
\end{eqnarray}

\noindent
On special choice of  $\Gamma$:
\begin{eqnarray}
c_{3}=0: \; \Longrightarrow \;  -\cos {\Gamma \over 2 }  u_{4} +
\sin {\Gamma \over 2}   u_{3} =0  ,
\label{B.16a}
\end{eqnarray}

\noindent that  is
\begin{eqnarray}
\tan {\Gamma \over 2} = \; {u_{4} \over u_{3} } \; ,
\nonumber
\end{eqnarray}

\noindent
expressions for   $B(c)$   and $O(c)$ are  much simplified:
 \begin{eqnarray}
 c_{1} = \cos { \Gamma  \over 2}  \; {u_{2} u_{3} -u_{4} u_{3} \over u_{3} } \; ,
\qquad
 c_{2} =  \cos { \Gamma  \over 2}   \;\; { u_{1} u_{3} + u_{2} u_{4} \over u_{3} } \; ,
\nonumber
\\
c_{3}= 0 \; , \qquad
c_{4} = \cos { \Gamma  \over 2}   \; { u_{3}^{2}  + u_{4}^{2} \over u_{3} } \; ,
\label{B.17a}
\end{eqnarray}

\noindent 3-vector parameter is
\begin{eqnarray}
C_{1} = {c_{1} \over c_{4} } = {u_{2} u_{2} -u_{4} u_{1} \over u_{3}^{2} + u_{4}^{2}  } =
-\; {n_{2} \over 1 - n_{3}} \; ,
\nonumber
\\
C_{2} = {c_{2} \over c_{4} } =  \;\;  {u_{1} u_{3} +u_{2} u_{4} \over u_{3}^{2} + u_{4}^{2}  }
=
 {n_{1} \over 1 - n_{3}} \; ,
 \;\;
 C_{3} =0 \; .
\label{B.17b}
\end{eqnarray}

\noindent
this $\vec{C}$  is directed  along
\begin{eqnarray}
\vec{n} \times \vec{n}_{(-)} =
\left | \begin{array}{ccc}
\vec{e}_{1} & \vec{e}_{2} & \vec{e}_{3} \\
n_{1} & n_{2} & n_{3} \\
0 & 0 & -1
\end{array} \right | = -n_{2} \; \vec{e}_{1} + n_{1} \; \vec{e}_{2} \; .
\nonumber
\end{eqnarray}

\noindent
With
\begin{eqnarray}
n_{1} = \sin \theta \cos \phi  ,\;
n_{2}  = \sin \theta \; \sin \phi  , \;
n_{3} = \cos \theta \; ,
\nonumber
\end{eqnarray}

\noindent for modulus of the  vector  we get
\begin{eqnarray}
\mid \vec{C} \mid =  + \sqrt{{n_{1}^{2} +n_{2}^{2} \over (1 -n_{3})^{2}}} =
\tan \; ({\pi - \theta \over 2} ) \; .
\nonumber
\end{eqnarray}

GENERALIZATION
\vspace{5mm}

Let us find  $SU(2)$-rotation, transforming spinor  $\Psi$  into $\Psi'$:
\begin{eqnarray}
B(c) \; \Psi = \Psi ' \; ,
\label{B.19}
\end{eqnarray}

The  task  may be solve  by two ways. First, let us  consider eq.
(\ref{B.19}) in its real-valued form
(compare with (\ref{B.14b}))
\begin{eqnarray}
\left | \begin{array}{rrrr}
u_{4} & -u_{1}  & u_{2} & u_{3} \\
u_{1} & u_{4}  & -u_{3} & u_{2} \\
u_{2} & -u_{3}  & -u_{4} & -u_{1} \\
u_{3} & u_{2}  & u_{1} & -u_{4}
\end{array} \right |
\left | \begin{array}{c}
c_{4} \\
c_{1}\\
c_{2} \\
c_{3}
\end{array} \right |  =
\left | \begin{array}{c}
u_{4}' \\
u_{1}'\\
u_{2}' \\
u_{3}'
\end{array} \right |   ,
\label{B.20a}
\end{eqnarray}

\noindent its  solution is
\begin{eqnarray}
c_{4} = u_{4}' \; u_{4} + u_{1}' \; u_{1}   + u_{2}' \;u_{2} +
u_{3}'\;u_{3} \; ,\;\;\;
\nonumber
\\
c_{1} = u_{1}' \; u_{4} - u_{4}' \; u_{1}   - u_{2}' \; u_{3} +
u_{3}'\;u_{2} \; ,\;\;\;
\nonumber
\\
c_{2} = - u_{2}' \; u_{4} + u_{4}' \; u_{2}   + u_{3}'\; u_{1} -
u_{3}'\; u_{1} \; ,
\nonumber
\\
c_{3} = - u_{3}' \; u_{4} + u_{4}'\; u_{3}   + u_{1}'\; u_{2} -
u_{2}'\; u_{1} \; .
\label{B.20b}
\end{eqnarray}

Else one, independent,  more simple and  symmetric consideration
is  based on   describing of normalized spinor by
unitary $2\times2$ matrices, when action of SU(2)-rotation is  given by
\begin{eqnarray}
B(\hat{c}) \; B(u) = B(u')  \qquad \Longrightarrow
\nonumber
\\
B(\hat{c}) = B(u') \; B^{-1}(u) =  B(u') \; B(\bar{u})\; .
\label{B.21a}
\end{eqnarray}

Result (\ref{B.21a}) coincides with  (\ref{B.20b}); indeed,  eq.(\ref{B.20b})
is written as
\begin{eqnarray}
\hat{c}_{4} = u_{4}' \; \bar{u}_{4} - u_{1}' \; \bar{u}_{1}   -
u_{2}' \;\bar{u}_{2} - u_{3}'\; \bar{u}_{3} \; ,
\nonumber
\\
\hat{c}_{1} = u_{1}' \; \bar{u}_{4} + u_{4}' \; \bar{u}_{1}   +
u_{2}' \; \bar{u}_{3} -
 u_{3}'\;\bar{u}_{2} \; ,
\nonumber
\\
\hat{c}_{2} =  u_{2}' \; \bar{u}_{4} + u_{4}' \; \bar{u}_{2}   +
u_{3}'\; \bar{u}_{1} - u_{3}'\; \bar{u}_{1} \; ,
\nonumber
\\
\hat{c}_{3} =  u_{3}' \; \bar{u}_{4} + u_{4}'\; \bar{u}_{3}   +
u_{1}'\; \bar{u}_{2} -
 u_{2}'\; \bar{u}_{1} \; ,
\label{B.21b}
\end{eqnarray}

\noindent or
\begin{eqnarray}
\left. \begin{array}{l} \hat{c}_{4} = u_{4}' \; \bar{u}_{4} -
u_{i}' \; \bar{u}_{i}  \; ,  \qquad
\hat{c}_{i} = u_{i}' \; \bar{u}_{4} + u_{4}' \; \bar{u}_{i}   +
\epsilon_{ijk} u_{j}' \; \bar{u}_{k}\; .
\end{array} \right.
\label{B.21c}
\end{eqnarray}

One principal point should  be  noted: each of spinor equations
\begin{eqnarray}
B(c) \; \Psi = \pm \; \Psi  \; ,
 \label{B.22a}
 \end{eqnarray}

\noindent has only trivial solution; indeed, eqs.~(\ref{B.20b}) give
\begin{eqnarray}
c_{4} = \pm  ( u_{4} \; u_{4} + u_{1} \; u_{1}   + u_{2}
\;u_{2} + u_{3}\;u_{3} )
  =\pm \; 1 \; ,
\nonumber
\\
c_{1} = \pm ( u_{1} \; u_{4} - u_{4} \; u_{1}   - u_{2} \;
u_{3} + u_{3}\;u_{2} ) = \;0 \; ,
\nonumber
\\
c_{2} = \pm ( - u_{2} \; u_{4} + u_{4} \; u_{2}   + u_{3}\;
u_{1} - u_{3}\; u_{1} )
 = \; 0 \; ,
\nonumber
\\
c_{3} = \pm  ( - u_{3} \; u_{4} + u_{4}\; u_{3}   + u_{1}\;
u_{2} - u_{2}\; u_{1} ) = \;  0 \; ,
\nonumber
\end{eqnarray}

\noindent that is
\begin{eqnarray}
B(c) = \pm \; I \; .
\label{B.22c}
\end{eqnarray}

\begin{quotation}

{\em
In other words,  turning to spinor  description eliminates from the
formalism the  concept of
small group:
in the case of vectors it is isomorphic to $SO(2)$, in the case of
spinors it  is  reduced to trivial group of a single element, identity  $I$.}

\end{quotation}

\label{last}
\end{document}